\documentclass[a4paper,fleqn,usenatbib,useAMS]{rasti}

\usepackage{newtxtext,newtxmath}
\usepackage{amsmath}	
\usepackage{cleveref}
\Crefformat{equation}{Equation~#1}
\usepackage{graphicx}	
\usepackage{amsmath}	
\usepackage{multicol}        
\usepackage{bm}		
\usepackage{pdflscape}	
\usepackage[T1]{fontenc}
\usepackage{ae,aecompl}

\DeclareRobustCommand{\VAN}[3]{#2}
\let\VANthebibliography\thebibliography
\def\thebibliography{\DeclareRobustCommand{\VAN}[3]{##3}\VANthebibliography}

\newcommand{\fftvis}[0]{\texttt{fftvis}}
\newcommand{\matvis}[0]{\texttt{matvis}}
\newcommand{\finufft}[0]{\texttt{finufft} }

\title[FFT-Based Simulator for Radio Interferometry Arrays]{\fftvis: A Non-Uniform Fast Fourier Transform Based Interferometric Visibility Simulator}
\author[Cox et. al]{Tyler A. Cox$^{1,2}$\thanks{E-mail: tyler.a.cox@berkeley.edu}
Steven G. Murray$^{3}$,
Aaron R. Parsons$^{1,2}$,
Joshua S. Dillon$^{1,2}$,
Kartik Mandar$^4$,
\newauthor
Zachary E. Martinot$^{5}$,
Robert Pascua$^{6}$,
Piyanat Kittiwisit$^{7,8}$,
James E. Aguirre$^{5}$
\\$^{1}$Department of Astronomy, University of California, Berkeley, CA
\\$^{2}$Radio Astronomy Laboratory, University of California, Berkeley, CA
\\$^{3}$Scuola Normale Superiore, 56126 Pisa, PI, Italy
\\$^{4}$Department of Physics, Indian Institute of Science Education and Research Bhopal, Madhya Pradesh, India, 462066
\\$^{5}$Department of Physics and Astronomy, University of Pennsylvania, Philadelphia, PA
\\$^{6}$Department of Physics and Trottier Space Institute, McGill University, 3600 University Street, Montreal, QC H3A 2T8, Canada
\\$^{7}$Department of Physics and Astronomy, University of Western Cape, Cape Town, 7535, South Africa
\\$^{8}$ South African Radio Astronomy Observatory, Black River Park, 2 Fir Street, Observatory, Cape Town, 7925, South Africa
}
\date{Accepted 2025 November 12. Received 2025 November 10; in original form 2025 June 02
}
\pubyear{2025}
\graphicspath{{./}{figures/}}

\begin{document}
\label{firstpage}
\pagerange{\pageref{firstpage}--\pageref{lastpage}}
\maketitle

\begin{abstract}
The detection and characterization of the 21\,cm signal from the Epoch of Reionization (EoR) demands extraordinary precision in radio interferometric observations and analysis. For modern low-frequency arrays, achieving the dynamic range necessary to detect this signal requires simulation frameworks to validate analysis techniques and characterize systematic effects. However, the computational expense of direct visibility calculations grows rapidly with sky model complexity and array size, posing a potential bottleneck for scalable forward modeling. In this paper, we present \fftvis, a high-performance visibility simulator built on the Flatiron Non-Uniform Fast-Fourier Transform (\texttt{finufft}) algorithm. We show that \fftvis{} matches the well-validated \matvis{} simulator to near numerical precision while delivering substantial runtime reductions, up to two orders of magnitude for dense, many-element arrays. We provide a detailed description of the \fftvis{} algorithm and benchmark its computational performance, memory footprint, and numerical accuracy against \matvis, including a validation study against analytic solutions for diffuse sky models. We further assess the utility of \fftvis{} in validating 21\,cm analysis pipelines through a study of the dynamic range in simulated delay and fringe-rate spectra. Our results establish \fftvis{} as a fast, precise, and scalable simulation tool for 21\,cm cosmology experiments, enabling end-to-end validation of analysis pipelines.
\end{abstract}


\section{Introduction}
The detection and characterization of the Cosmic Dawn and Epoch of Reionization (EoR) through the redshifted 21\,cm hyperfine transition of neutral hydrogen represents one of the most promising, yet challenging frontiers in modern cosmology. These epochs mark key phases in the evolution of the universe, signaling the formation of the first stars and galaxies, and the emergence of large-scale structure. Observations of the 21\,cm line offer a unique probe of these early times, allowing insights into the astrophysics of reionization, the thermal and ionization history of the intergalactic medium (IGM), and the nature of primordial fluctuations. For reviews of the study of the EoR through the 21\,cm line, see \citealt{2006PhR...433..181F, 2012RPPh...75h6901P, 2020PASP..132f2001L}. 

Several ongoing and planned experiments aim to detect this faint signal, including the Murchison Widefield Array (MWA; \citealt{2013JPhCS.440a2033T, 2013PASA...30...31B}), the Low Frequency Array (LOFAR; \citealt{2013A&A...556A...2V, 2017ApJ...838...65P}), the Owens Valley Long Wavelength Array (OVRO-LWA; \citealt{2019AJ....158...84E}), the Hydrogen Epoch of Reionization Array (HERA; \citealt{2017PASP..129d5001D, 2024PASP..136d5002B}), and the Square Kilometer Array (SKA; \citealt{2015aska.confE..19S, 2015aska.confE...1K}). Prior work also includes the Giant Meter Wave Radio Telescope (GMRT; \citealt{2013MNRAS.433..639P}) and the Precision Array for Probing the Epoch of Reionization (PAPER; \citealt{2010AJ....139.1468P}). However, the 21\,cm signal is intrinsically faint, five orders of magnitude weaker in brightness temperature than the dominant astrophysical foregrounds such as Galactic synchrotron emission and extragalactic radio sources. This poses a formidable dynamic range challenge for both instrumentation and data analysis \citep{2020PASP..132f2001L}. Although no experiments have reported a detection of the 21\,cm power spectrum, many have put increasingly sensitive upper limits, allowing for rough constraints on the timing of heating in the early Universe \citep{2022ApJ...924...51A, 2022ApJ...925..221A, 2023ApJ...945..124H, 2025arXiv250305576M, 2025arXiv250500373G}.

A leading strategy for handling this foreground challenge is to exploit the so-called “EoR window”, a region of Fourier space relatively free from foreground contamination under idealized conditions \citep{2010ApJ...724..526D, 2012ApJ...756..165P, 2012ApJ...752..137M, 2012ApJ...757..101T, 2013ApJ...770..156H, 2014PhRvD..90b3018L}. The existence of this window relies on the assumption that astrophysical foregrounds are spectrally smooth, while the 21\,cm signal is expected to vary rapidly with frequency as it traces fluctuations in the ionization and thermal state of the intertergalactic medium across redshift. Accessing this window requires exquisitely precise calibration, accurate systematics modeling, and foreground mitigation techniques to prevent spectral leakage of foreground power into cosmological modes. The required level of accuracy is extreme---even minor inaccuracies in calibration, instrument modeling, or data processing can introduce frequency-dependent systematics that obscure the cosmological signal. This is especially consequential in interferometric measurements, where chromatic instrument responses couple spatial structure into spectral modes, contaminating the 21\,cm power spectrum. Foreground leakage induced by spurious spectral structure is one of the main obstacles to a detection of the 21\,cm signal. Overcoming this necessitates not only optimized hardware design, but also rigorous control of analysis pipelines, from calibration, systematics mitigation, and foreground modeling through to power spectrum estimation. 

As a result, accurate and efficient simulation tools have become an essential component of all modern 21\,cm experiments. By enabling controlled, end-to-end analyses, these tools provide insight into how specific choices impact the final power spectrum measurement. Such simulations allow for a comprehensive validation of specific components of an analysis pipeline, including understanding the influence of various instrumental systematics and their mitigation strategies \citep{2019ApJ...884..105K, 2022MNRAS.514.1804J, 2024arXiv240608549R}, the consequences of calibration errors \citep{2016MNRAS.461.3135B, 2017MNRAS.470.1849E, 2019ApJ...875...70B, 2019MNRAS.487..537O, 2020MNRAS.492.2017J}, the robustness of power spectrum estimators \citep{2018ApJ...868...26C}, the development and validation of new calibration frameworks \citep{2021MNRAS.503.2457B, 2022MNRAS.517..910S, 2022MNRAS.517..935S, 2022ApJ...938..151E, 2023ApJ...943..117B, 2024MNRAS.532.3375C}, and the characterization of the effects of radio frequency interference and incomplete data sampling \citep{2020MNRAS.498..265W, 2023MNRAS.520.5552P, 2024MNRAS.535..793B, 2025ApJ...979..191C}.

Additionally, the growing adoption of Bayesian inference methods in 21\,cm cosmology has highlighted an increasing interest in fast and precise visibility simulations \citep{2016ApJS..222....3Z, 2016MNRAS.462.3069S, 2017arXiv170103384S, 2019MNRAS.488.2904S, 2023MNRAS.520.4443B, 2023ApJS..266...23K, 2024MNRAS.534.2653M, 2024MNRAS.530.3412Z, 2024RASTI...3..607G, 2024RASTI...3..400W}. These Bayesian techniques typically focus on constraining individual aspects of the analysis, such as instrumental beam patterns, sky emission models, or systematics, utilizing forward modeling to constrain specific subsets of parameters efficiently. Recently, however, \citealt{2025arXiv250407090K} presented a forward modeling simulation framework that directly compares simulated visibilities against observed data, including a parameterized beam, sky, and 21\,cm model as part of the optimization routine, significantly increasing computational demands. This approach highlights an even more pressing need for computationally efficient visibility simulators, capable of supporting extensive parameter exploration within these complex, high-dimensional inference frameworks.

Perhaps most importantly, simulations provide a means to rigorously test and validate analysis pipelines from start to finish. By generating mock observations that incorporate the full complexity of real data, including realistic astrophysical and cosmological models, noise, and instrumental effects, experiments are now able to robustly assess signal loss, calibration bias, and foreground leakage by processing these simulations through to the final power spectrum estimation. Such end-to-end tests are essential for building confidence in upper limits reported \citep{2019PASA...36...26B, 2020MNRAS.493.1662M, 2022ApJ...924...85A, 2024PASA...41...67L}. Given the scale of modern radio arrays, such validation efforts often require simulating datasets of immense size and complexity, encompassing thousands of frequency channels and integrations, hundreds of antennas, and sky models with millions of components. As a result, advances in simulator efficiency and scalability are not only beneficial but are necessary for the continued progress and reliability of 21\,cm cosmological analyses.

Several existing simulators have been developed to meet these needs, including \texttt{pyuvsim} \citep{2019JOSS....4.1234L}, \texttt{healvis} \citep{2019ascl.soft07002L}, MeqTrees \citep{2010A&A...524A..61N}, Montblanc \citep{2015A&C....12...73P}, FHD \citep{2012ApJ...759...17S}, WODEN \citep{2022JOSS....7.3676L}, OSKAR \citep{2009wska.confE..31D}, \texttt{RIMEz} \citep{martinot.2022} and \texttt{hera\_sim}\footnote{\url{https://github.com/HERA-team/hera_sim}}. Each of these tools offers unique capabilities and trade-offs between computational efficiency, accuracy, and flexibility. For the HERA experiment, the matrix-based visibility simulator, \matvis{} \citep{2025RASTI...4....1K}, has become the simulator of choice for validation and explorative analysis tasks due to its combination of speed and high accuracy. By decomposing the direct evaluation of visibilities into an outer product of individual antenna responses, \matvis{} significantly improves simulation performance, especially for arrays with a large number of antennas. However, this approach becomes increasingly expensive when applied to the full HERA array, where the sheer number of antennas and the required sky model complexity push the technique's computational limits.

To address these scaling limitations for simulations of densely-packed, many-element arrays, we explore an alternative approach based on the non-uniform Fast Fourier Transform (NUFFT), which offers a route to accelerate visibility simulations by exploiting the Fourier structure of the radio interferometric measurement equation (RIME). In this paper, we introduce \fftvis, a radio interferometric visibility simulator that utilizes the Flatiron Institute’s NUFFT library, \texttt{finufft} \citep{barnett2019parallelnonuniformfastfourier}, for efficient evaluation of visibilities under the point source approximation. Building on the foundation laid by \matvis{}, \fftvis{} emphasizes computational scalability and memory efficiency, while simulating visibilities that match those generated by \matvis{} to near-machine precision. Whenever possible \fftvis{} also matches the existing application programming interface (API) of \matvis{} while offering speed-ups of up to \emph{multiple orders of magnitude} for simulations of compact, many-element arrays, making it an appealing alternative to \matvis{} in most situations where simulation speed is a bottleneck.

The remainder of this paper is organized as follows. In Section \ref{sec:algorithm}, we begin with a description of the core calculations made by visibility simulators, and detail the specific algorithm used by \fftvis. In Section \ref{sec:fftvis_package}, we detail the software aspects of \fftvis, which we make publicly available on \texttt{GitHub}. Section \ref{sec:validation} validates \fftvis{} by comparing its simulated visibilities against analytical visibility solutions of diffuse sky models. We also validate its utility as a simulator for 21\,cm cosmology applications by performing a study of the dynamic range achieved in the delay and fringe-rate spectra of simulated visibilities. In Section \ref{sec:performance_comparison}, we compare the computational and memory requirements of the \fftvis{} algorithm to the \matvis{} algorithm, including an evaluation of its multi-core scaling performance. We then conclude in Section \ref{sec:conclusion} with a summary of our findings and a discussion of future directions.

\section{Algorithm} \label{sec:algorithm}
In this section, we present the algorithm used by \fftvis{} for efficiently simulating radio interferometric visibilities. We begin by introducing the radio interferometer measurement equation (RIME) and discuss the computational challenges associated with its numerical evaluation. We then briefly review the various approximations and numerical techniques used in existing visibility simulators to address these challenges. Following this, we present an overview of the \finufft algorithm, highlighting how it enables an acceleration in the evaluation of simulated visibilities while maintaining a high degree of accuracy. Finally, we detail our approach to primary beam interpolation and coordinate transformations and summarize the key components of the algorithm.

\subsection{Evaluating the Measurement Equation}

The fundamental quantity in radio interferometric simulations is the visibility, $\mathbf{V}_{ij}$, which describes the measured interferometric response of a baseline formed by two antennas $i$ and $j$ to incident sky radiation. The Radio Interferometer Measurement Equation (RIME; \citealt{1996A&AS..117..137H, 2011A&A...527A.106S}) provides a general expression for this response, linking the observed visibilities to the sky brightness and the instrumental response via a sky integral modulated by a geometric fringe term. In its commonly used Jones-matrix formulation, the RIME for a single time-frequency snapshot is given by \citep{2019ApJ...882...58K}:

\begin{align}
\mathbf{V}_{ij}(\nu) = \int_{4\pi} \mathbf{A}_i(\nu, \hat{\mathbf{s}})\, \mathbf{C}(\nu, \hat{\mathbf{s}})\, \mathbf{A}_j^{\dagger}(\nu, \hat{\mathbf{s}})\, e^{-2\pi i \nu \mathbf{b}_{ij} \cdot \hat{\mathbf{s}} / c}\, d^2\Omega,
\label{eqn:RIME}
\end{align}
where $\hat{\mathbf{s}}$ is a unit vector on the celestial sphere, $\mathbf{b}_{ij}$ is the baseline vector between the two antennas, $c$ is the speed of light, and $\nu$ is the observing frequency. The antenna beam response, represented by matrix $\mathbf{A}_i(\nu, \hat{\mathbf{s}})$ represents the complex, polarized voltage response of antenna $i$ in the direction $\hat{\mathbf{s}}$ at frequency, $\nu$. The sky brightness is described by the coherency matrix $\mathbf{C}(\nu, \hat{\mathbf{s}})$, which, under the assumption of uncorrelated electric field components, takes the form:

\begin{equation}
\mathbf{C} = \begin{pmatrix}
    \langle E_\theta E^*_\theta \rangle &  \langle E_\theta E^*_\phi \rangle \\
    \langle E^*_\theta E_\phi \rangle & \langle E_\phi E^*_\phi \rangle
    \end{pmatrix} = 
    \begin{pmatrix}
    I + Q & U + iV \\
    U - i V & I - Q
    \end{pmatrix}.
\end{equation}
where $I$, $Q$, $U$, and $V$ are the usual Stokes parameters, and the matrix elements correspond to ensemble-averaged products of the orthogonal electric field components, $E_\theta$ and $E_\phi$, in the local spherical basis. For most simulations, $\mathbf{C}$ is assumed static in equatorial coordinates, reflecting a time-independent sky model. The product $\mathbf{A}_i \mathbf{C} \mathbf{A}_j^\dagger$ encodes the coupling of sky polarization into the visibility, as measured through the primary beam response of the two antennas. The exponential term in Equation \ref{eqn:RIME} introduces the direction-dependent geometric delay, modulating the contribution of each sky direction by a complex fringe pattern determined by the baseline vector. Though formally the integral is over the entire sky, in practice, the contribution is limited to regions within the antenna field of view and above the horizon at the time of observation.

Although the RIME offers a rigorous mathematical foundation for simulating interferometric visibilities, its direct analytic evaluation is generally impractical due to the complex and often poorly understood nature of antenna primary beams and the sky brightness distribution. As a result, all visibility simulators adopt approximations and numerical strategies to discretize or parameterize the sky, allowing the numerical evaluation of the RIME. A widely used approach is the point source approximation, wherein the sky is modeled as a set of discrete sources, each assigned a specific flux, and the RIME integral is correspondingly replaced by a summation. Alternative bases, such as spherical harmonics or other decompositions, are also used in some simulators (e.g. \texttt{RIMEz}) to compactly represent diffuse emission and enable semi-analytic treatments of the RIME, particularly on the full sky.

To accommodate the computational demands and modeling accuracy required by modern 21\,cm experiments, a wide range of visibility simulators have emerged, each striking a different balance between accuracy, speed, and flexibility depending on its target application. For example, \texttt{pyuvsim} prioritizes per-baseline RIME evaluations with minimal approximations, making it well-suited for precision validation studies at the cost of computational efficiency. Others, such as OSKAR and WODEN, achieve greater computational throughput by leveraging GPU acceleration, but are built around simulating the SKA-Low and MWA phased-array station. More recently, \matvis{} introduced a matrix-based formulation that preserves the mathematical structure of RIME while significantly reducing runtime through efficient antenna-based matrix operations, allowing accurate visibility simulations with antenna-level scaling, a key advantage for large-$N$ arrays. However, as noted earlier, \matvis{} still encounters scalability bottlenecks when applied to many-element arrays or sky models with large numbers of sources.

The \fftvis{} simulator, introduced in this work, builds on the foundation established by \matvis{} but introduces alternative algorithmic optimizations targeting improved simulation performance on densely-packed, many-element arrays. Rather than operating in antenna space, \fftvis{} recasts the visibility equation as a sum of complex exponentials evaluated via a Type-3 Non-Uniform Fast Fourier Transform (NUFFT). This approach exploits the Fourier structure inherent in the measurement equation and enables the simultaneous computation of visibilities across all baselines from a single NUFFT call per time and frequency. The result is a simulator that matches the numerical precision of \matvis{} while achieving order-of-magnitude improvements in runtime and memory usage, particularly for compact, densely-populated arrays. 

The core \texttt{fftvis} algorithm proceeds frequency-by-frequency iterating over time as an inner loop, operating on user inputs including antenna positions, a sky model (specified as equatorial source coordinates and Stokes I intensities), and an antenna primary beam model. For each time step and frequency channel, it performs the following operations:
\begin{enumerate}
    \item \textbf{Coordinate Transformation:} Equatorial source positions from the input sky model are rotated into the topocentric horizontal frame using coordinate transformation routines adapted directly from the \matvis{} codebase. The available transformation schemes and their associated trade-offs are summarized in Section~\ref{subsec:coordinate_rotation}.

    \item \textbf{Beam Application:} The primary beam response (assuming a single beam model for all antennas in the current implementation) is evaluated or interpolated at the rotated source positions and applied to the source intensities to determine complex source amplitudes $c_j$ (detailed in Section~\ref{subsec:beam_interpolation}).

    \item \textbf{NUFFT Visibility Calculation:} The core visibility summation over all $N_{\rm sources}$ is efficiently computed for all $N_{\rm bls}$ baselines simultaneously using a Type-3 Non-Uniform Fast Fourier Transform (NUFFT), using the \texttt{finufft} library. This is the key innovation in this paper, allowing us to avoid the direct $O(N_{\rm sources} N_{\rm bls})$ summation or formation of large matrices. The specifics of the NUFFT algorithm and its implementation are detailed in Section~\ref{subsec:finufft}.

    \item \textbf{Output Assembly:} The visibilities $V_{ij}(\nu, t)$ are assembled for all baselines and written to output arrays compatible with \texttt{hera\_sim} or other downstream tools.
\end{enumerate}
This NUFFT-based reformulation significantly accelerates the visibility evaluation compared to direct summation or traditional matrix methods for many relevant simulation scenarios, as discussed in more detail in in Section~\ref{sec:performance_comparison}.

\subsection{Non-Uniform Fast-Fourier Transform Evaluation of RIME}
\label{subsec:finufft}

\begin{figure*}
    \centering    \includegraphics[width=\textwidth]{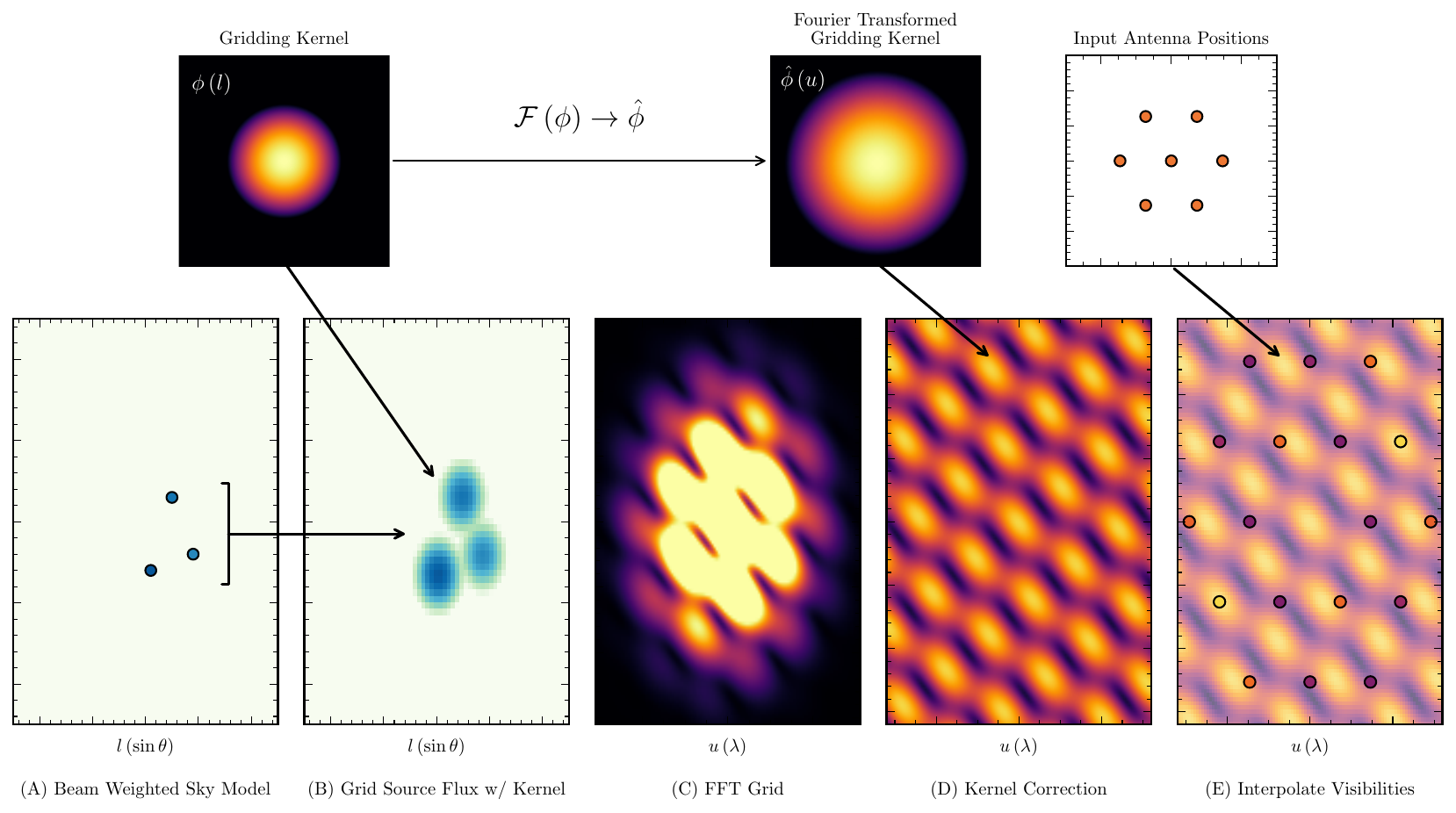}
    \caption{Schematic representation of the Flatiron Institute Non-Uniform Fast Fourier Transform (\texttt{finufft}) algorithm as utilized by \fftvis{}. The simulation pipeline consists of five key stages: (\textbf{Panel A}) The user-supplied sky model is weighted by the antenna beam pattern, which is interpolated to each source component's position. (\textbf{Panel B}) The beam-weighted source intensities are projected onto a fine grid through convolution with a compact gridding kernel $\phi$. (\textbf{Panel C}) The gridded sky model undergoes transformation from image-domain to visibility-domain via Fast Fourier Transform (FFT). (\textbf{Panel D}) The visibility-domain grid is corrected for gridding artifacts through deconvolution with the Fourier transform of the gridding kernel, $\hat{\phi}$. (\textbf{Panel E}) Finally, the algorithm interpolates the gridded visibilities to the exact sampling positions in the $uv$-plane determined by the unique baseline vectors derived from the input antenna positions. This process enables efficient computation of visibilities from arbitrary source distributions and antenna positions while maintaining numerical accuracy within user-specified NUFFT tolerance parameters.}
    \label{fig:pedagogy_plot}
\end{figure*}

Traditional numerical evaluation of the measurement equation involves directly summing complex exponentials for each baseline-source combination, an approach leading to prohibitive computational complexity, especially for many-element arrays or large sky catalogs. To accelerate the simulation of visibilities under the point source approximation, \fftvis{} utilizes non-uniform Fast Fourier transforms (NUFFTs). By reformulating the RIME as a Fourier-like summation over discretized sky positions, \fftvis{} utilizes the high-performance Type 3 NUFFT implementation provided by the \texttt{finufft} library \footnote{The non-uniform Fast Fourier Transform generalizes the FFT to allow for non-uniform sampling in the input or output domains. The Type 1 NUFFT maps non-uniform spatial data to uniform frequency modes, Type 2 maps uniform spatial data to non-uniform frequencies, and Type 3 transforms between arbitrary non-uniform sources and target points. See \citet{barnett2019parallelnonuniformfastfourier} for full definitions.} \citep{barnett2019parallelnonuniformfastfourier}, which enables fast and accurate visibility computations, particularly advantageous for dense interferometric arrays.

Under the point-source approximation, the visibility integral is approximated as
\begin{equation}
V(\nu) \approx \sum_{j=1}^{N_{\rm sources}} c_j \, e^{-2\pi i \nu \mathbf{b} \cdot \mathbf{\hat{s}}_j/c},
\end{equation}
where $\mathbf{s}_j$ represents the non-uniform sky positions of $N_{\rm sources}$ point sources, and $c_j$ denotes an element of the $2 \times 2$ matrix product $\mathbf{A}^\dagger \mathbf{C} \mathbf{A}$.
This summation is structurally equivalent to a Type 3 NUFFT, also known as the "non-uniform to non-uniform" transform. In its general form, the Type 3 NUFFT aims to efficiently evaluate the outputs $f_k$ at $N$ arbitrary target Fourier modes $\mathbf{x}_k \in \mathbb{R}^d$ given $N_{\rm sources}$ non-uniform source points $\mathbf{s}_j \in \mathbb{R}^d$ with complex strengths $c_j \in \mathbb{C}$ as follows:
\begin{equation}
f_k := \sum_{j=1}^{N_{\rm sources}} c_j e^{i \mathbf{x}_k \cdot \mathbf{s}_j}, \quad k = 1, \ldots, N_{\rm bls}. \label{eq:type3_nufft}
\end{equation}
In the context of visibility simulation, the target modes $\mathbf{x}_k = -2\pi \nu \mathbf{b}_k / c$ correspond to baseline coordinates scaled by frequency and therefore vary non-uniformly with both baseline vector ($\mathbf{b}$) and frequency ($\nu$). The source positions $\mathbf{s}_j$ are also non-uniformly distributed on the sky. A direct evaluation of the sum in Equation~\ref{eq:type3_nufft} requires $\mathcal{O}(N_{\rm sources} N_{\rm bls})$ operations, which becomes computationally expensive for the large numbers of sources ($N_{\rm sources}$) and baselines ($N_{\rm bls}$) typical in wide-field visibility simulations.

The \texttt{finufft} algorithm evaluates this sum extremely efficiently via a three-stage process:
\begin{enumerate}
    \item \textbf{Spreading (Gridding) Step}: First, the complex source amplitudes, $c_j$, located at their non-uniform positions $\{\mathbf{s}_j\}$, are effectively "spread" onto a temporary, intermediate uniform grid. This step involves convolving the sparse source data with a compact, smooth kernel function. \texttt{finufft} utilizes a kernel based on the "exponential of semicircle" function,
    \[
    \phi(z) = e^{\beta \left(\sqrt{1 - z^2} - 1\right)},
    \]
    defined on the interval $[-1, 1]$ where $\beta = 2.3 w$ is a scalar factor proportional to the size of the gridding kernel. This kernel is selected for its favorable properties, including smoothness and rapid decay in the Fourier domain. The spatial width of the convolution kernel balances the accuracy of the subsequent FFT against the computational cost of the gridding operation. This is achieved by defining a half-width parameter $w$ that is related to the requested precision $\epsilon$, typically via $w = \lceil \log_{10}(1/\epsilon)\rceil + 1$, where $\lceil \cdot \rceil$ is the ceiling operator. This convolution step maps the non-uniformly sampled source data onto a regular grid for efficient processing via FFT, while the kernel's properties help minimize errors introduced during this gridding and subsequent interpolation. The size of this intermediate grid ($N_{\rm grid}$) is also chosen based on $\epsilon$ and the source point and target Fourier mode extents.

    \item \textbf{Oversampled FFT}: After gridding the source terms, a standard Fast Fourier Transform (FFT) is performed on the intermediate, oversampled grid. The oversampling ensures that Fourier-domain aliasing artifacts, introduced by the preceding gridding step, remain below the requested precision level. This FFT efficiently transforms the gridded spatial information into the visibility domain.
    
    \item \textbf{Deconvolution and Interpolation}: Finally, the desired visibility values $f_k$ corresponding to the specific non-uniform target baseline coordinates $\{\mathbf{x}_k\}$ are extracted from the transformed grid resulting from the FFT. This involves two sub-steps: (a) \textit{Deconvolution}: The distorting effect of the spreading kernel convolution applied in Step 1 is corrected by dividing the FFT output by the analytic Fourier transform of the gridding kernel. (b) \textit{Interpolation}: The deconvolved spectrum, still defined on the uniform intermediate grid, is then interpolated to the exact target coordinates $\{\mathbf{x}_k\}$ corresponding to the requested baseline vectors and frequencies, yielding the final NUFFT output.
\end{enumerate}

The three-stage evaluation of the Type 3 NUFFT (i.e., spreading, FFT, and interpolation) is formally analogous to the “gridding and FFT” procedure traditionally used in interferometric imaging (e.g., \citealt{2017isra.book.....T}). In both approaches, irregularly sampled data are convolved with a compact kernel to map them onto a uniform grid, transformed by a Fast Fourier Transform, and then interpolated back to the desired coordinates. The distinction lies primarily in the choice and interpretation of the convolution kernel. In \texttt{finufft}, the spreading kernel is selected for its numerical properties -- compact support, smoothness, and an analytically tractable Fourier transform -- that jointly minimize aliasing in the Fourier domain and enable efficient, vectorized evaluation. The “exponential of semicircle” kernel adopted by \texttt{finufft} provides near-optimal accuracy for a given kernel width while maintaining a small computational footprint, which is the primary source of the algorithm’s efficiency \citep{barnett2019parallelnonuniformfastfourier}.

Conceptually, one could instead adopt a physically motivated convolution kernel, such as the Fourier transform of the instrument’s primary beam, to incorporate beam-weighting directly in Fourier space. This choice would make the NUFFT’s gridding operation equivalent to beam-weighted imaging, potentially compact in the $uv$-space for wide-field instruments. However, substituting such a kernel would generally sacrifice the analytic and separable form that enables the high precision and performance of the optimized NUFFT kernel. While the Type 3 NUFFT shares its theoretical foundation with traditional imaging, its efficiency arises from the deliberate use of a mathematically optimized, rather than physically derived, gridding kernel.

A subtle but essential step in the Type 3 NUFFT is selecting the size of the intermediate uniform grid ($n_i$) used for the Fast Fourier Transform stage. Unlike Type 1 and Type 2 transforms, where the grid size is directly related to the number of input or output points, the Type 3 grid must be sufficiently fine to accurately represent the transformation between the potentially wide-ranging non-uniform input source positions $\{\mathbf{s}_j\}$ and the non-uniform output baseline coordinates $\{\mathbf{x}_k\}$. Specifically, to control aliasing and interpolation errors introduced during the gridding and interpolation steps, this intermediate grid must possess adequate resolution across the full range spanned by both the spatial structure of the sources and the Fourier structure determined by the baselines. For each dimension $i = 1, \dots, d$, \texttt{finufft} chooses the grid size $n_i$ according to the following:
\begin{equation}
n_i = \left\lceil \frac{2 \sigma}{\pi}X_i S_i + w \right\rceil
\label{eqn:grid_size}
\end{equation}
where $S_i = \max_j |s^{(i)}_j|$ represents the maximum extent of the source coordinates in dimension $i$, and $X_i = \max_k |X^{(i)}_k|$ represents the maximum extent of the target baseline coordinates in units of wavelengths in that dimension. The parameter $\sigma > 1$ is a grid oversampling factor (typically $\sigma = 2$) used to mitigate aliasing effects. The product of these terms $X_i S_i$ captures the core requirement of the NUFFT: a larger extent in either the spatial domain ($S_i$) or the Fourier domain ($X_i$) requires a larger number of grid points ($n_i$) to maintain accuracy. 

It is important to note that the overall intermediate grid size, $N_{\rm grid} = \prod_i^d n_i$ (defined more formally below), is thus driven by these maximum extents, $X_i$ and $S_i$, rather than by the sheer number of sources ($N_{\rm sources}$) or baselines ($N_{\rm bls}$) contained within those extents. Effectively, the grid must be fine enough to resolve the highest Fourier variations ($X_i$) across the full spatial extent ($S_i$) of the input, plus an additional margin determined by the kernel-width ($w$) and oversampling factor ($\sigma$) to achieve the target precision ($\epsilon$). While larger $X_i S_i$ products increase the grid size and thus the cost of the FFT, \texttt{finufft} mitigates this by automatically translating both $\{\mathbf{x}_j\}$ and $\{\mathbf{s}_k\}$ to be centered around zero, minimizing $X_i$ and $S_i$ and hence reducing the grid size. However, when simulating wide-field visibilities (large $S_i$) or including baselines with large $|\mathbf{b}|/\lambda$ (large $X_i$), the required grid size can still become substantial. Conversely, if either the angular extent of the sky being simulated is small (small $S_i$), or the array's extent in wavelengths is limited (small $X_i$), $N_{\rm grid}$ can remain more manageable, which is beneficial for NUFFT performance regardless of $N_{\rm sources}$ or $N_{\rm bls}$ within those compact extents. In such cases, the performance may depend sensitively on the geometry of the problem.

The precise form of this scaling behavior is described by the computational complexity of the Type 3 transform in \texttt{finufft}, which scales as
\begin{equation}
    O\left(N_{\rm sources} w^d + N_{\rm grid} \log N_{\rm grid} + N_{\rm bls} w^d\right),
\end{equation}
where $d$ is the dimensionality of the input array (typically $2$ or $3$, depending on whether the input antenna coordinates are 2- or 3-dimensional) and $N_{\rm grid}$ is the size of the oversampled, intermediate grid, defined as
\begin{equation}
    N_{\rm grid} = \prod_i^d n_i
\end{equation}
This computational scaling reflects the three-stage evaluation of the RIME, in which $N_{\rm sources}$ point sources are convolved onto the intermediate grid with a kernel of size $w^d$, then Fourier transformed via an FFT, which has a computational scaling of $\mathcal{O}\left(N_{\rm grid} \log N_{\rm grid}\right)$, and finally interpolated to $N_{\rm bls}$ baselines with the Fourier transform of the gridding kernel, which also has size $w^d$. This provides a substantial speedup over a direct $O(N_{\rm sources} N_{\rm bls})$ evaluation for simulation cases where the number of baselines or the number of sources in the sky model is large. The NUFFT advantage over direct summation is particularly maximized when the intermediate FFT grid size ($N_{\rm grid}$) remains manageable. This occurs not only when the physical array size in units of wavelengths is small (small $X_i$), but also when simulating a compact region of the sky (small $S_i$), as $N_{\rm grid}$ depends on these extents rather than directly on $N_{\rm sources}$ or $N_{\rm bls}$. However, most high redshift 21\,cm applications require a full-sky, horizon-to-horizon sky model. In this case, we can write the scaling of the grid size in more familiar terms,
\begin{equation}
n_i = \left\lceil 4 \sigma \frac{b_{i, \rm max} \nu}{c} + w \right\rceil.
\label{eqn:grid_size_full_sky}
\end{equation}
A schematic summary of the \texttt{finufft} algorithm as used by \fftvis{} can be found in Figure~\ref{fig:pedagogy_plot}.

\subsubsection{Evaluating the Measurement Equation for Gridded Arrays}
\label{subsubsec:type1}
While the Type 3 NUFFT provides a general method for visibility simulation, further computational speedups can be achieved if the array's physical baseline vectors, $\mathbf{b}$, exhibit specific regularity. For instance, if the set of all $\mathbf{b}$ vectors inherently forms a regular Cartesian grid, meaning each baseline component $b_x, b_y, b_z$ is an integer multiple of a fundamental physical spacing $\Delta b_x, \Delta b_y, \Delta b_z$ respectively, this array configuration becomes an ideal candidate for Type 1 NUFFT application. This transform computes values from non-uniform source positions $\mathbf{x}_j$ onto a uniform grid of output modes. In the case of an array with Cartesian-gridded baselines, the desired visibilities at these specific $\mathbf{b}$ locations are directly represented by the values on this uniform output grid, after accounting for the $\nu/c$ scaling factor that relates physical baseline lengths to the output modes at the observing frequency $\nu$.

The Type 1 NUFFT can also be adapted for arrays where their antenna positions form a regular but non-Cartesian lattice (e.g. a hexagonal grid). In these instances, a linear transformation $\mathbf{L}$ is chosen to map the physical baselines $\mathbf{b}_{\rm phys}$ into a set of gridded baselines, $\mathbf{b}_{\rm gridded} = \mathbf{L}\mathbf{b}_{\rm phys}$, which are, by design, arranged on a regular Cartesian grid. To ensure that the phase term ($-2\pi i \nu \mathbf{b}_{\rm phys} \cdot \mathbf{s}_j/c$) in the visibility sum is accurately computed, the original source coordinates $\mathbf{s}_j$ must also be transformed to $\mathbf{s}_j' = (\mathbf{L}^T)^{-1}\mathbf{s}_j$. The Type 1 NUFFT then calculates the visibilities values corresponding to these gridded baseline configurations $\mathbf{b}_{\rm gridded}$ using the transformed source coordinates $\mathbf{s}_j'$ as input.

The efficiency of this Type 1 NUFFT strategy is best understood through its computational scaling. Let $N_{\rm unif}$ represent the total number of points in the uniform output grid targeted by the Type 1 NUFFT where this $N_{\rm unif}$ is determined by the extent and desired resolution of the set of gridded baseline configurations (either natively Cartesian $\mathbf{b}$ or transformed $\mathbf{b}_{\rm gridded}$). The complexity for this Type 1 transform scales as 
\begin{equation}
    \mathcal{O}(N_{\rm sources} w^d + N_{\rm unif} \log N_{\rm unif}).
\end{equation} 
This can offer a significant performance advantage over the general Type 3 NUFFT's $\mathcal{O}(N_{\rm sources} w^d + N_{\rm grid} \log N_{\rm grid} + N_{\rm bls} w^d)$ scaling, particularly if $N_{\rm unif}$ is substantially smaller than $N_{\rm grid}$ (the Type 3 intermediate grid size from Equation~\ref{eqn:grid_size}). This occurs when the array is regular, but very sparse. $N_{\rm grid}$'s size is strongly influenced by the product of the source extent $S_i$ and the maximum extent of the target modes $X_i$ (where $X_i$ relates to the maximum baseline components scaled by $\nu/c$). In contrast, $N_{\rm unif}$ reflects the number of points needed to represent the specific set of gridded baseline configurations of the array. If this count is smaller than $N_{\rm grid}$ (especially if $S_i$ is large, which tends to inflate $N_{\rm grid}$ for Type 3), the Type 1 NUFFT provides a more efficient route for simulating visibilities.

By default, the \fftvis{} codebase automatically inspects the user-supplied antenna positions to determine whether they lie on a lattice, and therefore whether a computationally cheaper Type 1 NUFFT can be substituted for the general Type 3 case.  This procedure computes the set of unique baseline separations in each Cartesian direction and evaluates whether these separations can be expressed as integer multiples of a common fundamental spacing within a user-defined numerical tolerance. If this condition is satisfied, \fftvis{} identifies the corresponding lattice vectors, constructs the linear transformation matrix $\mathbf{L}$ that maps the physical baseline vectors $\mathbf{b}_{\rm phys}$ onto a regular grid ($\mathbf{b}_{\rm gridded} = \mathbf{L}\,\mathbf{b}_{\rm phys}$), and applies the appropriate inverse transpose $(\mathbf{L}^T)^{-1}$ to the input sky coordinates.  The transformation is fully automated, and no requires no manual pre-processing of antenna or source coordinates by the user. When the array geometry does not meet the lattice criteria, either because the spacings are irregular beyond the specified tolerance or the configuration is inherently non-gridded, \fftvis{} defaults to the Type 3 NUFFT. This ensures numerical correctness for arbitrary array layouts while exploiting the Type 1 speedup whenever possible.  

\subsubsection{Considerations for Non-Flat Arrays}

The NUFFT framework and its performance characteristics, including the determination of $N_{\rm grid}$ (Equation \ref{eqn:grid_size}) and the overall scaling involving $w^d$, apply generally to $d$ dimensions. For arrays that are effectively co-planar, or where baseline $w$-components (the component parallel to the pointing direction) are negligible, simulations can often utilize a 2-dimensional NUFFT ($d=2$) which is typically less computationally intensive. However, when an array contains antenna with significant non-coplanar positions, its baseline vectors $\mathbf{b} / \lambda=(u,v,w)$ will include substantial $w$-components. To accurately model the source-baseline geometry in these scenarios, the dot product $\nu \mathbf{b} \cdot \mathbf{s}_j/c$ in the visibility equation must be treated in its full 3-dimensional form.

Using a 3D NUFFT (that is, setting $d=3$ in the Type 3 transform algorithm as used by \fftvis{}) has significant consequences for the computational cost, impacting each stage of the NUFFT process detailed earlier. Firstly, the gridding of $N_{\rm sources}$ onto the intermediate grid and the final interpolation from this grid to $N_{\rm bls}$ baselines now involve a 3-dimensional convolution kernel. This means that their operational costs, which scale with $w^d$, will now scale with $w^3$ rather than $w^2$. Secondly, and often more importantly, the size of the intermediate FFT grid, $N_{\rm grid}$, becomes a 3D product, $N_{\rm grid} = n_x n_y n_z$. Each $n_i$ (for $i \in \{x,y,z\}$) is determined by Equation~\ref{eqn:grid_size}, so $n_z$ will depend on the maximum extent of the product of the baseline $w$-components (contributing to $X_z$) and the corresponding source coordinate ($S_z$, related to the $n_j-1$ term in direction cosines). Even if $n_z$ (the grid dimension along the $w$-axis) is smaller than $n_x$ or $n_y$, its inclusion substantially increases the total number of points in $N_{\rm grid}$ compared to a 2D grid of $n_x n_y$. As a result, the FFT stage, with complexity $\mathcal{O}(N_{\rm grid} \log N_{\rm grid})$, becomes much more computationally intensive due to this significantly larger $N_{\rm grid}$.

The impact of these increased operational costs, from the 3D gridding kernel and especially the larger 3D FFT grid, is that simulating visibilities for non-flat arrays using a 3D NUFFT is inherently more computationally expensive than an analogous 2D NUFFT simulation for a flat array of similar $u,v$ extent and source count. This can reduce the overall speedup factor that the NUFFT provides compared to direct summation methods. Therefore, the 3-dimensional nature of an array and the necessity of a 3D NUFFT represent an important consideration when evaluating expected simulation performance and selecting the most appropriate algorithmic strategies for visibility generation.

\subsubsection{Assumption of Identical Antenna Beams}
\label{subsubsec:fftvis_beams}
In the simplified scenario where all antennas share a common beam model, the coefficients $c_j$ are independent of baseline and can be precomputed uniformly across the array for each time step and frequency channel. This uniformity enables all visibilities at a given time and frequency to be evaluated through a single Type-3 NUFFT call, substantially reducing the computational overhead involved in visibility simulations.  In contrast, when antenna beams vary across the array, the direction-dependent response becomes inherently baseline-specific. Under these more general conditions, each unique pair of antenna beams defines a distinct interferometric response to the sky. As a result, the number of required independent NUFFT evaluations per time-frequency snapshot increases to
\begin{equation} 
N_{\rm NUFFT} = \frac{N_{\rm beam} (N_{\rm beam} + 1)}{2} \end{equation} 
where $N_{\rm beam}$ is the number of unique beam patterns present within the array. If we assume that the NUFFT step of the simulation dominates the total runtime, we expect this will increase the simulation time by a factor of $N_{\rm NUFFT}$. In contrast, the matrix-based approach used by \matvis{} has a more favorable scaling with $N_{\rm beams}$ since the core visibility computation depends on the number of antennas rather than the number of unique beam patterns. To maintain computational efficiency, the current implementation of \fftvis{} assumes the beam model supplied by the user is shared by all antennas. While this constraint remains well-justified for many practical simulation scenarios, certain validation and analysis tasks require support for per-antenna beams \citep{2019MNRAS.487..537O, 2024RASTI...3..400W, 2022ApJ...941..207K, 2023ApJ...953..136K}. Future releases of \fftvis{} will include support for heterogeneous, antenna-dependent beam configurations.

\subsection{Coordinate Rotation}
\label{subsec:coordinate_rotation}

Accurate visibility simulation demands careful treatment of celestial coordinate transformations, particularly the mapping between equatorial coordinates, where sky models are typically defined, and the topocentric East-North-Up (ENU) frame, which is more natural for expressing instrumental quantities such as baseline vectors and primary beams. Small inaccuracies in these coordinate transformations can propagate into significant phase errors, especially for long baselines or high-frequency observations, where sub-arcsecond misalignments correspond to considerable visibility decorrelation.

\fftvis{} adopts the coordinate transformation strategy introduced in \matvis{} \texttt{v1.3.0}. Specifically, \fftvis{} can utilize one of two approaches, including (i) full astrometric corrections using \texttt{astropy} at each time-step or (ii) a hybrid approach that applies computationally intensive corrections such as precession, nutation, and aberration less frequently (potentially only once at the start of the simulation) rather than at every simulation time step.

By default, \fftvis{} uses the hybrid method where full astrometric corrections via the \texttt{astropy} library are applied only periodically (at user-specified intervals) to pre-correct source coordinates. Between these corrections, computationally efficient rigid body rotations, dependent on local sidereal time and latitude, transform these pre-corrected equatorial coordinates to the horizontal (ENU) frame at each time step. In \citealt{2025RASTI...4....1K}, the authors demonstrate that this hybrid transformation approach reproduces the results of full-precision coordinate evaluations with excellent accuracy for snapshot observations. 

However, because the transformation matrix $\mathbf{R}(t)$ is constructed assuming a fixed celestial sphere, errors due to neglected time-dependent astrometric effects (e.g., precession, proper motion, and nutation) increase with time since the reference epoch. Specifically, slow-varying effects like precession and nutation are applied only once during the \texttt{astropy}-based pre-correction at the reference time, not during the per-timestep rigid-body rotations. Because these astrometric effects change very slowly, their error contribution is negligible for observations taken at or near this reference time. These simplifications result in periodic errors in the simulated visibilities on a 24-hour sidereal cycle relative to the high-precision simulations produced by \texttt{pyuvsim}, which utilizes \texttt{astropy} for its coordinate rotations and transformations. The error is quantified in \citealt{2025RASTI...4....1K}, where the authors show that for snapshot observations, the hybrid method matches high-precision simulations with fractional residuals of $10^{-10}$.

From a simulation standpoint, the primary advantage of this scheme lies in its computational efficiency: once equatorial positions are pre-corrected, subsequent evaluations require only matrix multiplications that are easily vectorized and parallelized. However, this same simplification limits its long-term astrometric accuracy, particularly for simulations requiring precision modeling for calibration purposes. For many 21\,cm cosmology use cases, especially those involving short observations, the hybrid method achieves a practical balance between computational cost and astrometric accuracy. For longer simulations requiring higher precision, \citealt{2025RASTI...4....1K} recommend mitigating this error by splitting the observation into shorter chunks and refreshing the full astrometric correction at a regular interval. In their analysis, the authors found that updating these astrometric corrections every $\sim 15$ minutes of simulation time kept fractional visibility errors to less than $\sim 10^{-5}$. As this interval is user-specified in \fftvis{}, the $\sim 15$ minute recommendation from the \matvis{} paper provides guidance on how to determine the necessary update interval to balance computational cost and the required astrometric accuracy.

For use cases demanding full astrometric rigor -- particularly for calibration tasks -- \fftvis{} provides the flexibility to use full \texttt{astropy}-based transformations to be performed at every simulation time step. Although this option ensures higher positional accuracy, it also incurs a significantly higher computational cost, potentially becoming a dominant simulation bottleneck. This flexibility in choosing the coordinate transformation method, inherited from \matvis{}, allows users to tailor the simulation performance and accuracy to their specific scientific requirements.

\subsection{Beam Interpolation}
\label{subsec:beam_interpolation}

Another component in achieving accurate simulations is the precise representation of antenna beam patterns evaluated at continuously evolving sky coordinates. This requirement presents particular challenges as sources drift across the field of view, necessitating beam evaluation at arbitrary sky coordinates where empirical beam measurements or simulations may not have explicitly been covered. Our beam interpolation strategy follows the well-established techniques used by the \matvis{} codebase backed by \texttt{pyuvdata} \citep{2017JOSS....2..140H, 2025JOSS...10.7482K}, ensuring that our approach takes advantage of thoroughly validated computational methods for beam handling and interpolation. The framework accommodates two principal classes of antenna beam models:
\begin{enumerate}
\item \textbf{Analytic Beam Models:} These models allow closed-form mathematical expressions that enable direct evaluation at arbitrary coordinates and frequencies without intermediate interpolation steps, maximizing computational efficiency and precision.
\item \textbf{Numerical Beam Models:} These discretely sampled representations, typically derived from electromagnetic simulations such as CST or HFSS, require sophisticated interpolation strategies to determine beam response values at arbitrary sky positions.
\end{enumerate}

For beam data management and interpolation operations, we integrate the \texttt{AnalyticBeam} and \texttt{UVBeam} classes from the \texttt{pyuvdata} package, maintained by the Radio Astronomy Software Group (RASG)\footnote{https://radioastronomysoftwaregroup.github.io/}. This framework supports various beam representation formats, including regularly gridded beam models, HEALPix-formatted beam distributions, and CST-simulated electromagnetic beam patterns. The \texttt{UVBeam} class provides multiple interpolation functions with configurable order parameters, including bilinear, bicubic, and quintic interpolation schemes operating on altitude-azimuth coordinate systems. These methods use the computational interfaces \texttt{scipy.interpolate.RectBivariateSpline} and \texttt{scipy.ndimage.map\_coordinates} to achieve high-precision beam evaluation at arbitrary source positions.

There are a few practical considerations, emphasized in \citealt{2025RASTI...4....1K}, that become especially important in high-accuracy simulations. One concerns the choice of interpolation scheme: when modeling visibilities for calibration or foreground subtraction, where sub-percent level deviations can become important, higher-order interpolation methods are recommended to maintain the accuracy of the beam response. However, the authors did note that for validation purposes, bilinear interpolation did not introduce appreciable systematic effects in the 21\,cm power spectrum estimate, indicating that lower-order interpolation schemes are sufficient for applications that do not demand full physical realism. Another relates to the coordinate system used for beam definition. Many E-field beam models exhibit phase discontinuities at the zenith in at least one polarization component. To handle these correctly, the beam should be defined on a spherical grid centered at the zenith, allowing interpolation routines to trace the field structure smoothly and avoid introducing artifacts associated with non-polar or Cartesian grids.

\subsection{Caveats and Limitations}

The \fftvis{} algorithm, as detailed in this section, provides an efficient NUFFT-based framework for simulating radio interferometric visibilities. The current implementation prioritizes performance for common 21\,cm cosmology use cases, particularly drift-scan observations with identical antenna beams. While powerful within this scope, we highlight several areas that present opportunities for future development to broaden its applicability.

As previously discussed in \ref{subsubsec:fftvis_beams}, \fftvis{} assumes a single, shared beam model across all antennas. This constraint is a result of the poor scaling of \fftvis{} with the number of unique antenna patterns. The current version of \fftvis{} limits the simulations to a single antenna beam that is shared across the array. This decision was made to preserve the computational efficiency and minimize the memory usage of the simulator. Extending support to per-antenna beam models is algorithmically straightforward but would introduce non-trivial performance costs when simulating arrays with many different beam types, particularly during beam evaluation and NUFFT stages. We defer this to future development. 

Additionally, the current version of \fftvis{} is optimized for drift-scan observations, as is typical for instruments like HERA and OVRO-LWA. This design choice aligns with the observing strategies of many 21\,cm experiments, where the sky drifts across a fixed primary beam and the instrument passively records visibilities. Simulating tracking observations, where the phase center follows a fixed sky position, requires additional complexity: not only must visibilities be rephased to a moving pointing center, but the primary beam response must also be updated over time to reflect the changing orientation of the array relative to a source. These features are compatible with the underlying formalism and may be incorporated in future releases as simulation use cases broaden beyond drift-scan mode.

These areas define the main directions for future development of \fftvis{}. However, the current implementation, focused on speed and efficiency for large-N arrays and sky models under common assumptions, provides a robust and validated tool for many simulation tasks in 21\,cm cosmology, as we demonstrate in the following sections.

\section{The \fftvis{} Package} \label{sec:fftvis_package}

The core functionality of the \fftvis{} Python package is encapsulated in the \texttt{fftvis.simulate} module, which provides a high-level API to simulate visibilities across a range of arbitrary sky models, frequency ranges, and time integrations. When feasible, the API design mirrors that of \matvis, allowing users to switch between the two simulators with minimal changes to their simulation infrastructure.

The primary interface, \texttt{fftvis.simulate\_vis}, accepts a sky model (as either a list of point sources or a HEALPix map), an array configuration, a beam model (supporting both analytic and gridded forms via the \texttt{pyuvdata.BeamInterface}), and metadata specifying the observing cadence and frequency channels. Internally, the simulator rotates the sky to topocentric coordinates at each time step using \texttt{numpy} matrix multiplication functions, evaluates the beam response with \texttt{scipy} interpolation routines, and grids the product onto the Fourier plane using \texttt{finufft}’s Type 3 NUFFT algorithm. The coordinate rotation code used in \fftvis{} is adapted directly from the \matvis{} codebase, ensuring consistent and efficient transformations and reducing the possibility of mismatched assumptions between simulators. This design choice further allows for the interoperability between the two tools. Parallelization over frequencies and time integrations is achieved via the \texttt{ray} library \citep{2017arXiv171205889M} and native OpenMP threading within \texttt{finufft}. 

For convenient integration into existing simulation workflows, \fftvis{} is also accessible through the \texttt{visibilities} module of the \texttt{hera\_sim} package, a simulation suite tailored to HERA and similar redundant arrays, maintained by the HERA collaboration. This integration supports configuration via \texttt{pyuvsim}-style YAML files and offers a unified interface that enables \fftvis{} to serve as a drop-in replacement for other visibility simulators, including \matvis{} and \texttt{pyuvsim}, without requiring extensive reconfiguration. 

\section{Simulator Validation} \label{sec:validation}

\begin{figure*}
    \centering
\includegraphics[width=\textwidth]{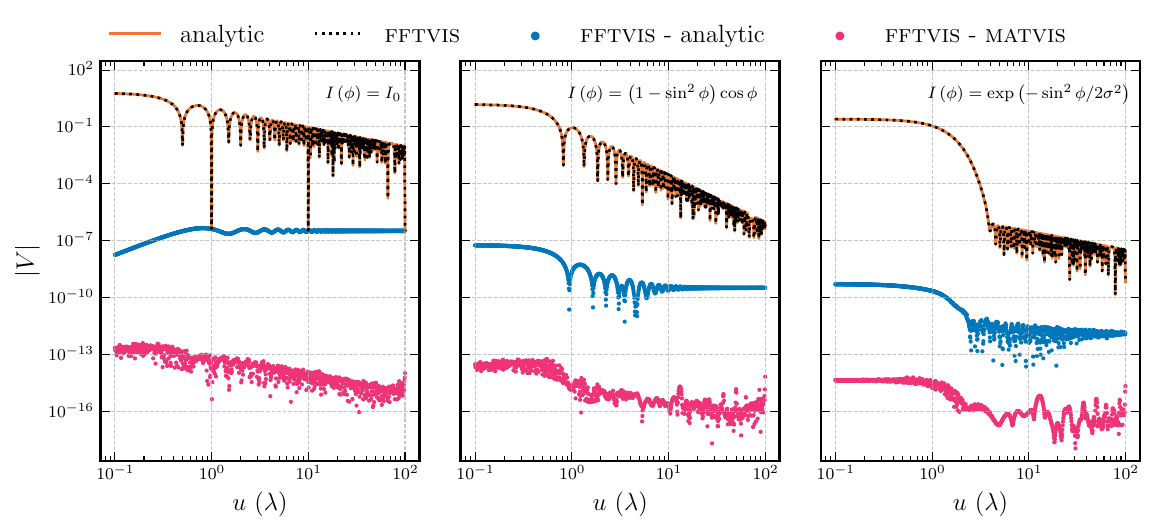}
    \caption{Validation of \fftvis{} against analytic full-sky visibility solutions for diffuse emission patterns from \citet{2022ApJS..259...22L}. Each panel shows the magnitude of the visibilities $|V|$ for a different sky models varying in zenith angle (top curves), along with absolute value of the difference $|V_1 - V_2|$ (bottom curves) between \fftvis{} and the analytic result (blue), and between \fftvis{} and \matvis{} (pink). All simulations were performed using discretized sky models generated by the \texttt{analytic\_diffuse} package. \fftvis{} matches the analytic solutions to better than $\sim$$10^{-5}$ for most of the range of $u$ values simulated, and agrees with \matvis{} to near machine precision, confirming the accuracy and reliability of the NUFFT-based method for simulating smooth, diffuse sky structure. We note that this sky model error is the result of HEALPix gridding errors, and will decrease with the size of the input sky model.}
    \label{fig:analytic_diffuse}
\end{figure*}

In this section, we describe our approach to validating \fftvis{} for use in 21\,cm simulation applications. Our strategy focuses on two complementary tests that together assess both the absolute numerical accuracy of the simulator and its suitability as a tool for validating 21\,cm data analysis pipelines.

The first test focuses on evaluating the intrinsic numerical accuracy of \fftvis{} by comparing its outputs against analytic visibility solutions derived from simplified diffuse sky models. These models admit closed-form expressions for the visibilities, allowing for an objective and implementation-independent benchmark \citep{2022ApJS..259...22L}. The agreement with these solutions provides a strong validation of the accuracy of the mathematical and numerical implementation of the simulation. The second test assesses \fftvis{} in the context of its intended use as a validator for 21\,cm analysis pipelines. Specifically, we examine the spectral and temporal coherence of the simulated visibilities by computing their delay and fringe-rate spectra and comparing with the visibilities produced by the \matvis{} simulator. With foregrounds exceeding the cosmological signal by 4–6 orders of magnitude, even subtle spectral artifacts from simulations can introduce spurious structure that contaminates the EoR window. To be useful as a validation tool, the simulator must avoid such artifacts and preserve the expected spectral smoothness of the simulated visibilities. This test examines whether \fftvis{} achieves the dynamic range and fidelity necessary to support robust pipeline validation and foreground modeling. For all validation tests presented in this section, we use the following software versions: \matvis{}\texttt{=1.2.1}, \texttt{pyuvsim=3.0.0}, and \fftvis{}\texttt{=1.0.0}. Unless otherwise stated, we set the NUFFT precision to $\epsilon = 10^{-13}$ for all tests.

\subsection{Comparisons to Analytic Diffuse Models}

Accurate modeling of the full-sky interferometric response to diffuse emission is an essential test for validating visibility simulators, particularly in foreground-dominated applications such as 21\,cm cosmology. Although point sources offer a simpler computational path via direct summation, accurately modeling diffuse emission requires robust numerical integration of the RIME across the sky, often by making approximations to decompose the diffuse model. Verifying a simulator's ability to correctly capture these features is essential to build confidence in its results.

\citet{2022ApJS..259...22L} developed a suite of analytically tractable
sky brightness distributions designed to test and validate the numerical accuracy of visibility simulators under realistic wide-field conditions. These test patterns are constructed to model the structural and angular characteristics of typical astrophysical emission (i.e., horizon-spanning profiles, sharp horizon cuts, etc.), while remaining mathematically simple enough to permit either closed-form or rapidly converging series solutions to the RIME. These analytic solutions provide an absolute reference against which the fundamental accuracy of a simulator's implementation can be assessed. For our analysis, we consider three test cases:
\begin{enumerate}
    \item A pure constant profile, $I(\phi)=I_0$, (\texttt{MONOPOLE} model, Section 3.1.1), which models horizon-spanning emission profiles;
    \item A modulated cosine, $I(\phi)=(1-\sin^{2}\phi)\cos\phi$, (\texttt{QUADDOME} model, Section 3.1.2),
    \item A Gaussian lobe ($I(\phi)=\exp(-\sin^{2}\phi/2\sigma^{2})$), (\texttt{GAUSS} model, Section 3.2.2), modeling compact, beam-shaped emission. For the comparison shown here, we adopt $\sigma=0.5$ corresponding to the "medium" \texttt{GAUSS} pattern.
\end{enumerate}
These models are discretized onto a full-sky HEALPix grid \citep{2005ApJ...622..759G} with $N_{\mathrm{side}}=1024$ using the \texttt{analytic\_diffuse} package\footnote{https://github.com/aelanman/analytic\_diffuse}, and fed into both \fftvis{} and \matvis{} for evaluation. Including \matvis{} in this comparison serves two key purposes. First, it provides a direct cross-check against an established simulator that has been independently validated. Second, since both \fftvis{} and \matvis{} operate on the same discretized HEALPix input in this test, comparing their outputs helps isolate any differences arising specifically from the core visibility calculation method (i.e., \fftvis{}'s NUFFT approach versus \matvis{}'s matrix-based direct summation) from the errors inherent in the point-source approximation used for the input sky model.

Figure~\ref{fig:analytic_diffuse} presents the results of this validation, showing the amplitude of the visibilities along with the absolute residuals for each of the three sky models. Across all test cases, \fftvis{} simulated visibilities show good agreement with the analytic solutions, with absolute errors typically below $10^{-5}$. Additionally, \fftvis{} matches \matvis{} to nearly machine precision for all models simulated. This indicates that the NUFFT algorithm, as implemented in \fftvis{}, accurately reproduces the results of the direct summation method used by \matvis{} when given identical inputs.

The remaining discrepancies between the simulator outputs (for both \fftvis{} and \matvis{}) and the analytic solutions arise primarily not from errors in the visibility calculation algorithms themselves, but from the approximations inherent in using a finite HEALPix grid to represent a continuous diffuse sky. As
discussed in \citealt{2022ApJS..259...22L}, the point source approximation, used by both \matvis{} and \fftvis{}, leads to integration errors, particularly near the horizon where geometric delays are largest, the fringe term oscillates most rapidly across a pixel, and sharp cut-offs in sky brightness or beam response occur. Even small misrepresentations in the boundaries of the pixels can result in significant amplitude deviations at high $u$. These effects are inherent to any simulator using the point source approximation on pixelized maps, and diminish with increasing angular resolution (i.e., higher $N_{\mathrm{side}}$). The excellent agreement between \fftvis{} and \matvis{} confirms that the NUFFT-based gridding used in \fftvis{} introduces no measurable bias relative to \matvis{}'s direct summation at the tested tolerance levels.

\subsection{Delay and Fringe-Rate Spectrum Dynamic Range}

\begin{figure*}
    \centering
    \includegraphics[width=\textwidth]{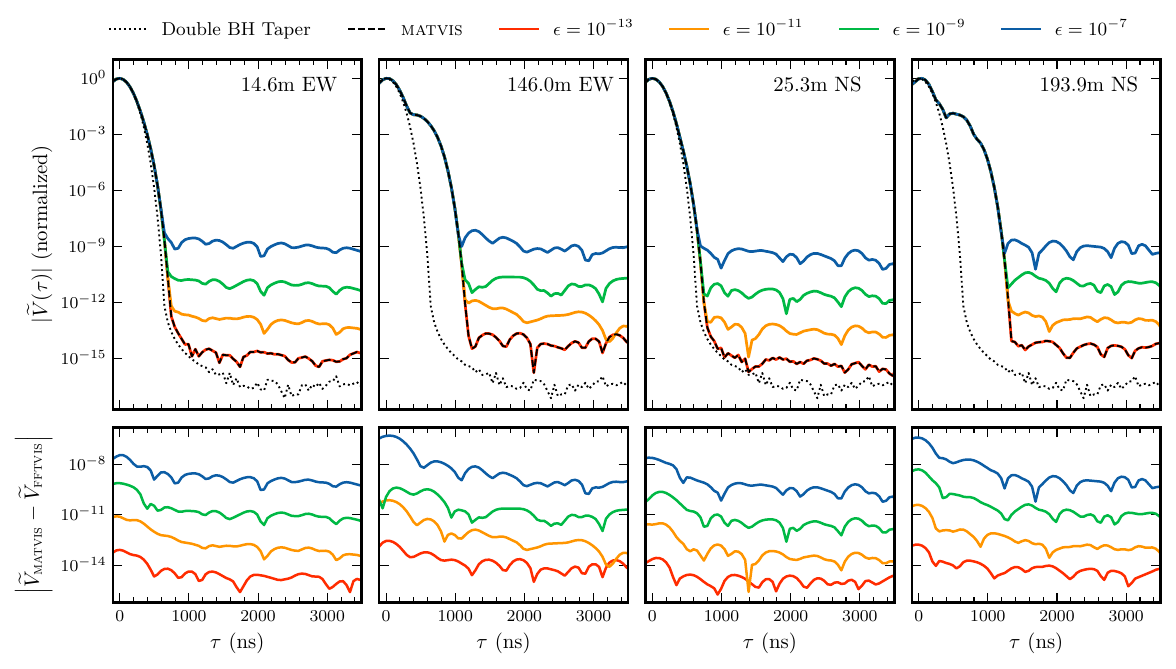}
    \caption{Delay spectra (\textbf{top row}) and the corresponding absolute difference between \matvis{} and \fftvis{} (\textbf{bottom row}) are shown for visibilities simulated using an airy beam model and smooth spectrum sky model for four different baselines (14.6m EW, 146.0m EW, 25.3m NS, and 193.9m NS). Here, we compare simulated visibilities from \matvis{} (dashed black), treated here as an exact numerical reference, and \fftvis{} at various NUFFT precision levels (colored lines). For context, the FFT of the squared Blackman-Harris tapering function (dotted black) indicates the dynamic range limit imposed by the spectral windowing itself. In the top row, we show the normalized amplitude of the delay transform for each set of simulated visibilities, while in the bottom row reveals numerical artificats introdcued by the NUFFT approximation in \fftvis{}, computed as the absolute difference relative to the \matvis{} reference. As the NUFFT precision parameter, $\epsilon$, decreases, \fftvis{} converges more closely to \matvis{} across the full range of delays. As $\epsilon$ decreases from $10^{-7}$ to $10^{-13}$, the \fftvis{} solutions demonstrate progressively improved convergence to the reference set of \matvis{} simulations. Even at moderate precision settings, \fftvis{} maintains numerical accuracy compared to \matvis{} exceeding $10^{-6}$ relative to foreground amplitudes. These results validate that the NUFFT-based approach introduces negligible algorithmic artifacts at precision levels appropriate for validating 21\,cm analysis pipelines.}
    \label{fig:finufft_precision}
\end{figure*}

The reliability of end-to-end validation efforts in 21\,cm cosmology depends on the numerical stability of the simulated visibilities used as input. In particular, a visibility simulator must preserve the intrinsic spectral and temporal coherence of the sky signal to avoid introducing spurious structure that could be misinterpreted as a failure of the analysis pipeline itself. This requirement is especially important in the context of power spectrum estimation and foreground mitigation, where many methods rely on the assumption that foregrounds are spectrally smooth and temporally coherent in a way that is consistent with the baseline vectors and sky rotation.

A sensitive metric of such spurious spectral structure is the delay spectrum, obtained by Fourier transforming visibilities along the frequency axis, 
\begin{equation} 
V_{ij}(\tau) = \int T(\nu) V_{ij}(\nu) e^{2\pi i \nu \tau} d\nu, 
\label{eqn:delay_spectra}
\end{equation} 
where $T(\nu)$ is a spectral taper (typically a Blackman-Harris taper for HERA) applied to suppress spectral sidelobes \citep{2012ApJ...756..165P}. In this domain, spectrally smooth foregrounds are confined to low-delay modes, while the 21\,cm signal appears at higher delays. The delay-space dynamic range, defined as the ratio of power at low delays to residual power at high delays, provides a stringent test of the spectral accuracy of the simulator.

If a simulator introduces numerical structure at a level comparable to or exceeding this dynamic range, it can lead to biased power spectrum estimates or false conclusions about the analysis performance. To be suitable for pipeline validation, a simulator must suppress such artifacts below the level of the 21\,cm signal across the full range of relevant Fourier modes. In this subsection, we test the numerical precision of \fftvis{} to determine whether it meets this standard. Of particular concern is the NUFFT step, which interpolates across frequency-dependent Fourier grids to compute visibilities and may introduce subtle spectral artifacts. These can manifest as excess power at high delays, where the sky signal should be negligible. We aim to demonstrate that, when operated at appropriate internal precision, \fftvis{} preserves spectral smoothness at the dynamic range required for accurate modeling of both the foregrounds and the cosmological signal. 

To determine whether \fftvis{} meets this standard, we evaluate its behavior across a range of internal precision settings, specifically the NUFFT tolerance parameter ($\epsilon$) which controls the allowable interpolation error during gridding. Lower values of $\epsilon$ yield more accurate transforms at higher computational cost. By systematically varying $\epsilon$ and comparing against \matvis{}, which has been independently validated for use in 21\,cm studies, we assess both the accuracy of the simulator and the presence of any artificial spectral structure. This allows us to identify precision settings where \fftvis{} maintains sufficient accuracy for high-dynamic-range forward modeling.

Our simulations for these delay-spectrum tests use a realistic foreground sky model composed of diffuse Galactic synchrotron emission derived from the Global Sky Model (GSM; \citealt{2017MNRAS.464.3486Z}) combined with a population of unresolved extragalactic point sources based on the GLEAM source-count distribution \citep{2019PASA...36....4F}. The sky model is then pixelized onto a full-sky HEALPix grid of size $N_{\mathrm{side}} = 256$. This resolution was chosen to adequately sample the sky structure in the GSM and prevent significant fringe aliasing for the longest baselines considered in the simulation. We simulate observations for four representative baselines drawn from the HERA array configuration: two east-west (14.6\,m, 146.0\,m) and two north-south (25.3\,m, 193.9\,m), selected to sample a range of geometric delays and orientations. The instrument's primary beam is modeled as an ideal frequency-dependent Airy pattern corresponding to HERA's 14-meter dish diameter, and the simulated observatory location is fixed to that of HERA. Visibilities are computed across a 100–120 MHz band using 160 channels, corresponding to a channel resolution of approximately 122\,kHz.

Figure~\ref{fig:finufft_precision} presents the resulting delay spectra in visibilities simulated using \fftvis{} with varying NUFFT precision settings, with residuals relative to the \matvis{} results in the bottom row. These panels clearly demonstrate that as $\epsilon$ is decreased from $10^{-7}$ to $10^{-13}$, the \fftvis{} delay spectra uniformly converge toward the \matvis{} reference across all baselines and delay modes. At the highest precision tested, the residuals between \fftvis{} and \matvis{} fall below $10^{-13}$, indicating agreement near numerical precision limits. Even at moderate precisions ($\epsilon \leq 10^{-9}$), \fftvis{} maintains spectral dynamic ranges exceeding $10^{8}$. This level of accuracy is well above the threshold required to ensure that numerical noise does not contaminate the foreground-dominated region of the delay spectrum or leak significantly into the EoR window. These results confirm that the NUFFT-based simulation approach implemented in \fftvis{} does not introduce measurable artificial spectral structure when operated at suitable precision levels, confirming its suitability for high-dynamic-range applications. Based on these findings, we recommend using a precision parameter of at least $\epsilon \leq 10^{-11}$ to ensure numerical artificats are sufficiently suppressed.

While decreasing $\epsilon$ (i.e., tightening precision) improves accuracy, it correspondingly increases the computational cost of the NUFFT. This cost increase is primarily influenced by the NUFFT gridding kernel width, $w$, which scales approximately as $\log(1/\epsilon)$. Several internal \texttt{finufft} operations depend on $w$, including aspects of source gridding and the determination of the intermediate FFT grid size (as detailed in Section~\ref{subsec:finufft}), making their runtime contributions $\epsilon$-dependent. However, because $w$ grows only logarithmically with $1/\epsilon$, even substantial changes in the requested precision (e.g., from $\epsilon = 10^{-7}$ to $10^{-14}$, which roughly doubles $w$) lead to relatively modest increases in the runtime components that scale polynomially with $w$. For instance, a component scaling as $w^d$ would increase by a factor of approximately $2^d$ (e.g., a factor of 4 if $d=2$). For this reason, we set the default value of $\epsilon = 10^{-13}$ in \fftvis{} to prioritize delay-space dynamic range over a small increase in simulator efficiency. However, we set this as a tunable parameter in the \fftvis{} API, allowing the user to decrease their simulation run time if delay-space dynamic range is not a strict requirement of their simulated visibilities.

Beyond the spectral characteristics tested by delay spectra, accurately simulating the temporal evolution of visibilities is essential for validating end-to-end analysis pipelines. Many 21\,cm analysis techniques rely on the assumption that astrophysical signals evolve smoothly over time to distinguish them from temporally variable systematics \citep{2024arXiv240608549R, 2024arXiv241001872P, 2024MNRAS.535.3218G, 2024MNRAS.534.3349C}. Therefore, simulated visibilities must replicate the expected time structure of the sky signal without introducing spurious variations from the simulation process itself. As with the spectral axis, a potential concern is that the NUFFT gridding step in \fftvis{}, which interpolates in $uv$-space independently at each timestep, may introduce subtle time-dependent artifacts. These could arise as the source coordinates evolve with Earth rotation, especially during intervals of rapid beam-weighted sky change such as when bright sources set on the horizon.

\begin{figure}
\centering
    \includegraphics[width=\columnwidth]{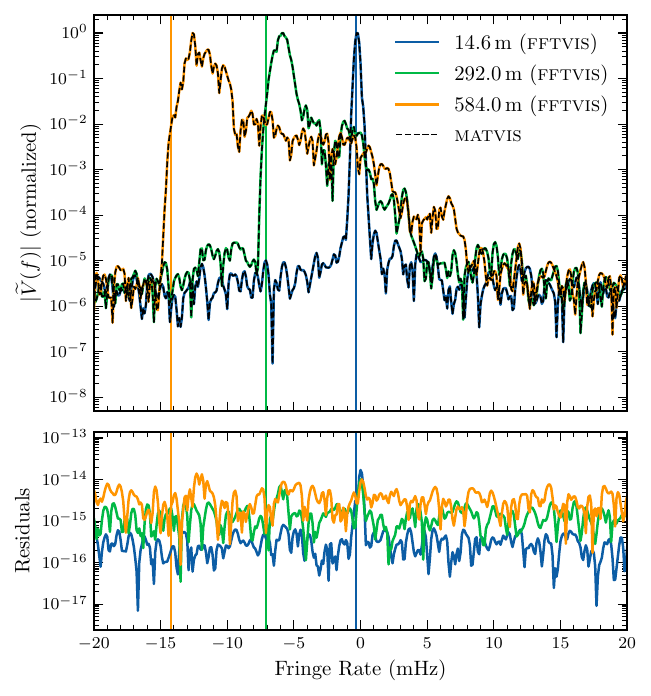}
    \caption{Fringe-rate spectra of simulated visibilities for three baselines with east-west projections of $14.6 \, \rm m$, $292.0 \, \rm m$, and $584.0 \, \rm m$, generated using \fftvis{} (using $\epsilon = 10^{-13}$) and \matvis{}. The top panel shows the normalized amplitude of the fringe-rate transformed visibilities, while the bottom panel presents the absolute residuals between the two simulators. Vertical dotted lines indicate the analytically expected maximum fringe rate, corresponding to the east-west projection associated with each baseline and latitude of the simulated array \citep{2016ApJ...820...51P} where the slight bleed outside of this window is caused by extra temporal structure introduced by sources moving through the beam. We find that the residuals between \fftvis{} and \matvis{} are below $10^{-13}$ for all baselines and fringe rates tested. The excellent agreement across all three baselines demonstrates that the use of the NUFFT in \fftvis{} does not introduce significant time-dependent artifacts into the simulated visibilities.}
    \label{fig:fringe_rate}
\end{figure}

To evaluate the temporal evolution of the visibilities, we simulated a long time series of 1000 integrations at 10-second intervals for baselines within the HERA array with a range of east-west projected lengths ($14.6$, $292.0$, and $584.0 \, \rm m$). We focus on Local Sidereal Times from 22.5 to 1.23 hours, an interval that includes the horizon crossing of the Galactic center, when sky-beam evolution is most rapid. From the resulting visibilities, we computed fringe-rate spectra by applying a tapered Fourier transform along the time axis. Figure~\ref{fig:fringe_rate} presents the resulting spectra generated from both \fftvis{} and \matvis{}, along with their absolute residuals. Across all baselines and fringe-rates tested, the simulators agree nearly to numerical precision, with residuals suppressed below $10^{-13}$. This result confirms that \fftvis{} accurately preserves the fringe-rate structure inherent to the simulated sky and observing configuration, introducing no measurable temporal artifacts even under conditions of rapid sky evolution.

In summary, these results demonstrate that \fftvis{}, when operated at high-precision settings, maintains both spectral and temporal coherence. It meets the numerical accuracy requirements necessary to support high-dynamic-range analyses in 21\,cm cosmology, including those that rely on delay-domain separation and fringe-rate filtering.

\section{Computational Considerations} 
\label{sec:performance_comparison}
In this section, we analyze the performance of \fftvis{} by comparing it to \matvis{}. We begin with an examination of the computational scaling of both \fftvis{} and \matvis{}, comparing their runtimes across various sky model sizes, antenna counts, and physical array dimensions. We also identify which steps of the \fftvis{} algorithm dominate its execution time, allowing targeted optimization in future development. Next, we focus on the memory usage of \fftvis{} by comparing the memory footprints of \fftvis{} and \matvis{}. Here, we highlight how \fftvis{} achieves substantially reduced memory requirements for many common simulation configurations by avoiding the creation of large intermediate matrices, in contrast to the \matvis{} approach. Finally, we explore the parallelization capabilities of \fftvis{} on multi-core architectures, demonstrating its ability to utilize parallel processing to further decrease simulation time.

Our timing tests were conducted at the NRAO New Mexico Array Science Center (NMASC)\footnote{\url{https://www.aoc.nrao.edu/computing/}}, on a full node with eight 2.6 GHz E5-2670 Xeon dual-core processors, with the number of active cores constrained at runtime for specific test cases. Individual runtime tests were measured using the \texttt{line\_profiler} Python package \citep{line_profiler}.

\subsection{Review of the \matvis{} algorithm}

In order to contextualize the performance comparison between \matvis{} and \fftvis{}, we briefly summarize the \matvis{} algorithm and its computational characteristics. The \matvis{} simulator reformulates the RIME in a matrix-based framework that exploits the separability of per-antenna and per-source quantities. Rather than directly evaluating the full per-baseline sum, an operation that scales as $\mathcal{O}(N_{\rm sources} N_{\rm bls})$, \matvis{} takes advantage of the correlation form of the measurement equation, in which a visibility can be expressed as a cross-correlation of antenna responses. This redefines the problem as $\mathcal{O}(N_{\rm sources} N_{\rm ants}^\alpha)$, where $\alpha \approx 1.3 - 2$ depending on data layout and caching efficiency, offering a substantial reduction in runtime for large arrays compared to direct evaluators of the measurement equation, such as \texttt{pyuvsim}.

Starting from the interferometric measurement equation, the geometric phase factor for source $n$ observed by the baseline between antennas $i$ and $j$ can be separated into per-antenna terms:
\begin{align}
e^{-2\pi i\, \mathbf{b}_{ij}\!\cdot\!\mathbf{y}_n(t)}
&= e^{-2\pi i\, \mathbf{x}_i\!\cdot\!\mathbf{y}_n(t)}\,
   e^{+2\pi i\, \mathbf{x}_j\!\cdot\!\mathbf{y}_n(t)} \nonumber\\
&\equiv \mathbf{F}_{in}\, \mathbf{F}_{jn}^*,
\end{align}
where $\mathbf{x}_i$ is the antenna position vector,  
$\mathbf{y}_n(t)$ encodes the direction cosine of source $n$ at time $t$,  
and $\mathbf{F}_{in}$ is the per-antenna phasor response.  The intrinsic source coherency matrix $\mathbf{C}_n$ (which couples Stokes parameters to the instrumental polarization basis) can be factored as
\begin{equation}
\mathbf{C}_n = \mathbf{M}_n\,\mathbf{M}_n^\dagger,
\end{equation}
where $\mathbf{M}_n$ is a Cholesky-like decomposition convenient for vectorized computation. Combining the direction-dependent primary beam response $\mathbf{A}_{ipkn}$ with the phasor and coherency matrices, the visibility for polarization products $p,q$ is
\begin{equation}
V^{pq}_{ij}(\nu_a,t)
= \sum_n
  \mathbf{A}_{ipkn}\,\mathbf{F}_{in}\,
  \mathbf{M}_{kk''n}\,
  \mathbf{M}_{k'k''n}^*\,
  \mathbf{F}_{jn}^*\,
  \mathbf{A}_{jqk'n}^* .
\end{equation}
To exploit matrix algebra, the per-antenna response to the entire sky is collected into an intermediate matrix
\begin{equation}
\mathbf{Z}_{ip,\,n k''}
= \sum_k
  \mathbf{A}_{ipkn}\,
  \mathbf{F}_{in}\,
  \mathbf{M}_{kk''n},
\end{equation}
which has shape $(N_{\rm ant}\!\times\!N_{\rm src})$ per polarization.  
The full visibility matrix is then obtained through an outer product:
\begin{equation}
V^{pq}_{ij}(\nu_a,t)
= \left[\mathbf{Z}\,\mathbf{Z}^\dagger\right]_{ij}^{pq}.
\label{eqn:matvis_mult}
\end{equation}
In this formulation, each column of $\mathbf{Z}$ encodes the complex beam-weighted sky response of a single antenna across all sources, while the matrix product $\mathbf{Z}\mathbf{Z}^\dagger$ efficiently computes all cross-correlations simultaneously.  
This converts the explicit per-baseline summation into a pair of dense matrix multiplications, achieving substantial reuse of per-antenna terms at the cost of storing the full $\mathbf{Z}$ matrix in memory whose size scales as $\mathcal{O}(N_{\rm sources} N_{\rm ants})$. 
\subsection{Runtime}
\label{subsec:runtime}

\begin{figure*}
    \centering    \includegraphics[width=\textwidth]{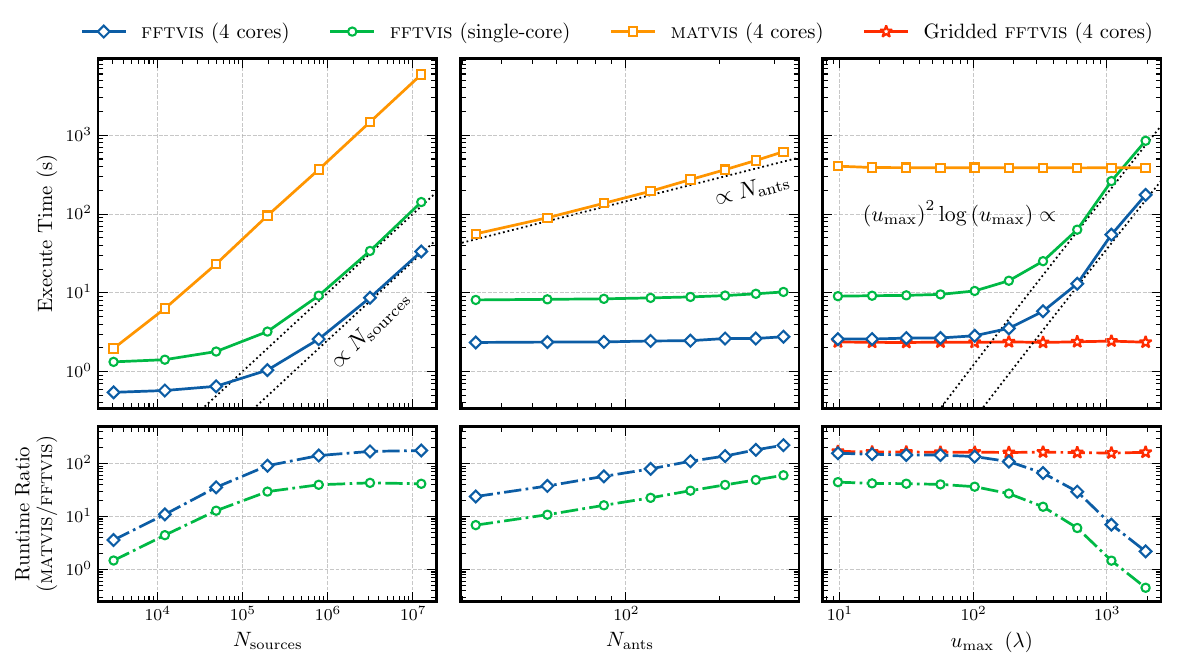}
    \caption{Here, we demonstrate how execution times (\textbf{top row}) of \matvis{} and \fftvis{} scale with three key simulation parameters: (\textbf{left column}) the number of sources in the sky model, (\textbf{middle}) the number of antennas in a densely packed array, and (\textbf{right}) maximum baseline length in wavelengths. We begin from a fiducial simulation that uses a HEALPix sky model of $N_{\rm side} = 256 \ (\approx 7.8 \times 10^5 \ \rm pixels)$, a densely-packaged hexagonal array of $N_{\rm ants} = 261$, and 32 time integrations at a single frequency channel. In each plot, we vary one parameter while holding the other parameters fixed. In the bottom row, we show the ratio of execution times (\matvis{}/\fftvis{}) for both single-core and 4-core CPU runs. From these panels, we find that \fftvis{} excels when simulating many-element, dense arrays, but struggles when the input array becomes more sparse in $uv$-coordinates, due to an under-utilization of the FFT of the input sky. However, an important exception occurs when antenna positions form a grid. In this scenario, a Type 1 NUFFT enables more efficient visibility simulation, as shown in the right-hand column.}
    \label{fig:scaling}
\end{figure*}

\begin{figure}
    \centering
    \includegraphics[width=\columnwidth]{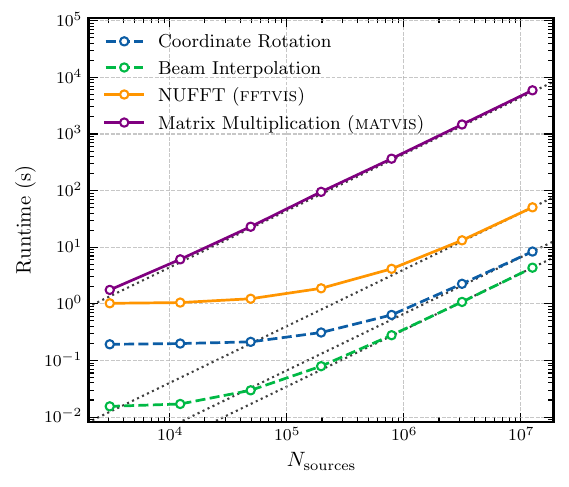}
    \caption{Breakdown of the relative contributions of the dominant sources of runtime as a function of the number of sources used for simulation for both \matvis{} and \fftvis{}. Here the computational steps that are shared between \fftvis{} and \matvis{} (coordinate rotation and beam interpolation) are marked with dashed colored lines while steps specific to each software package are marked by solid lines. As expected, the \texttt{finufft} stage dominates the total runtime for \fftvis{}, but is only a factor of a few times more than coordinate rotation and beam interpolation, and scales at the same rate (linearly with the number of sources). For \matvis{}, the total time is dominated by the final matrix multiplication step in Equation \ref{eqn:matvis_mult} and is typically many times more expensive than the coordinate rotation and beam interpolation stages of the algorithm.}
    \label{fig:algo_breakdown}
\end{figure}

To evaluate the computational efficiency of \fftvis, we performed a series of runtime scaling tests, mirroring the approach taken to characterize the performance of \matvis. Our primary goal was to understand how the execution time of \fftvis{} scales with key simulation parameters and to identify potential performance bottlenecks within the algorithm. These tests also help provide a reference for the cases where \fftvis{} may be the more efficient simulator than \matvis{} and vice versa.

Our fiducial set of simulation parameters consists of a HEALPix sky map with an $N_{\rm side} = 256 \ (N_{\rm sources} = 7.8 \times 10^5)$ and a HERA-like, split-core hexagonal array comprising 10 antennas per side ($N_{\rm ant} = 261$) simulated for 32 time integrations at a single frequency $(\nu = 100 \rm \ MHz)$. In our scaling experiments, we systematically vary three simulation parameters while keeping the other two parameters fixed at their default values to highlight the differences in computational scaling between the \fftvis{} and \matvis{} algorithms. Specifically, we examine the dependence of the runtime on the number of sources ($N_{\rm sources}$), the number of antennas ($N_{\rm ants}$), and the maximum baseline length ($u_{\rm max}$). For a consistent comparison between the simulators, we use the same coordinate rotation and beam interpolation schemes for both \matvis{} and \fftvis. Figure \ref{fig:scaling} summarizes the results of these tests. The top row shows the absolute runtime of both \matvis{} and \fftvis{} under single-core and 4-core CPU configurations. The bottom row displays the runtime ratio of \matvis{} to \fftvis{}, providing a direct measure of relative efficiency.

The left column of Figure~\ref{fig:scaling} demonstrates how the runtime of both simulators scales with the number of sources, $N_{\rm sources}$. Both \fftvis{} and \matvis{} exhibit approximately linear scaling, $\mathcal{O}(N_{\rm sources})$. This shared linear dependence can be partially attributed to operations performed per source component, primarily coordinate rotation and beam interpolation. While these steps contribute to the total execution time, the primary difference in performance is the result of the difference in the core visibility calculation methods. \matvis{} constructs and multiplies an antenna response matrix (computational cost scaling roughly as $\mathcal{O}(N_{\rm ant}^{\alpha} N_{\rm sources})$, where as discussed above, the factor $\alpha=1.3-2$ and depends on the efficiency of the matrix multiplication given the antenna/source matrix size), while \fftvis{} grids each source onto the Fourier plane for evaluation via a Type 3 NUFFT. 

As shown in Figure~\ref{fig:scaling}, the NUFFT-based gridding approach in \fftvis{} is substantially more efficient than the matrix multiplication used by \matvis{} when $N_{\rm sources}$ is large, leading to speedups exceeding two orders of magnitude for $N_{\rm sources} \gtrsim 10^6$ (using 4 CPU cores). A detailed breakdown of the internal runtime contributions within \fftvis{} as a function of $N_{\rm sources}$ is provided in Figure~\ref{fig:algo_breakdown}. This confirms that the NUFFT calculation itself, which also scales approximately linearly in this regime, remains the most expensive step. It is the efficiency of this NUFFT gridding relative to matrix multiplication that drives the observed performance advantage at high $N_{\rm sources}$.

In the middle column, we vary the number of antennas, $N_{\rm ants}$, arranged in a densely packed hexagonal configuration. While the number of baselines scales as $\mathcal{O}(N_{\rm ants}^2)$, the runtime of \matvis{} increases approximately linearly in this experiment, which is the result of the matrix-based approach taken by \matvis. \fftvis{}, on the other hand, exhibits a nearly flat scaling across this axis. Although its runtime does grow with the number of baselines, that dependence is subdominant to the cost of per-source operations such as coordinate rotation, beam interpolation, and gridding for the NUFFT. It is worth noting that in this simulation, $u_{\rm max}$ is held fixed, which means that increasing $N_{\rm ants}$ increases the density of the array, not its total size or angular resolution. As shown in the bottom row, the performance advantage of \fftvis{} continues to grow with increasing antenna count, reaching over two orders of magnitude for the largest configurations tested.

In the right column, we vary $u_{\rm max}$ by scaling the physical extent of the array while keeping $N_{\rm ants}$ and $N_{\rm sources}$ fixed. This increases the baseline lengths without changing the number of antennas or the sky model size. For \fftvis{}, increasing $u_{\rm max}$ requires a higher-resolution Fourier grid to resolve finer fringe structure and prevent aliasing, leading to a steep increase in runtime dominated by gridding and FFT operations. In contrast, \matvis{} shows relatively flat runtime in this experiment, as its cost is tied to the number of antenna-source pairs and is insensitive to baseline length when the sky resolution is held fixed. 

However, fixing the sky resolution is not realistic for simulations involving diffuse emission. In such cases, the angular resolution of the sky model must increase with $u_{\rm max}$ to accurately capture high-frequency fringe patterns and avoid aliasing, therefore the number of sources should sacle as $u^2_{\rm max}$. This requirement would increase the cost of \matvis, which then scales roughly as $\mathcal{O}(N^{\alpha}_{\rm ants} u^2_{\rm max})$. Meanwhile, the cost of \fftvis{} grows approximately as $\mathcal{O}(u^2_{\rm max} w^d + u_{\rm max}^2 \log u^2_{\rm max})$. Because of these scaling behaviors, \matvis{} is generally more efficient for sparse arrays with relatively few antennas, while \fftvis{} becomes the preferable choice for dense arrays, where the cost of computing visibilities for a large number of antennas becomes prohibitive for \matvis{}.

It is also important to note here that for arrays possessing a regular grid structure (either natively Cartesian or transformable to one, as detailed in Section~\ref{subsubsec:type1}), \fftvis{} has the capability to use a Type 1 NUFFT. Our timing results demonstrate that this approach is indeed much more efficient than the Type 3 transform for varying $u_{\rm max}$. This efficiency primarily arises from how the size of its uniform target grid ($N_{\rm unif}$ is determined. For a given gridded array, $N_{\rm unif}$ is set by the extent and resolution of the array's gridded baseline coordinates ($\mathbf{b}_{\rm gridded}$). Consequently, $N_{\rm unif}$ remains constant irrespective of the observing frequency $\nu$, making it independent of $u_{\rm max}$, as shown by the red line in Figure \ref{fig:scaling}. With both $N_{\rm sources}$ (fixed by the experiment) and $N_{\rm unif}$ (fixed by the array's physical baseline grid) being constant, the runtime for \fftvis{} using a Type 1 NUFFT becomes largely insensitive to changes in $u_{\rm max}$. This behavior contrasts sharply with the default Type 3 NUFFT, where the internal grid $N_{\rm grid}$ necessarily scales with $u_{\rm max}$ (via $X_i$ from Equation~\ref{eqn:grid_size}) and also with the sky extent $S_i$.

To provide a clearer picture of how these simulators compare under more general conditions, particularly considering how the number of sources ($N_{\rm sources}$) and maximum baseline extent ($u_{\rm max}$) influence runtime, we summarize their approximate high-level computational scalings. For \matvis{}, which performs a direct summation for each source-baseline pair, the computational cost is primarily determined by the product of the number of sources and a factor related to the number of antennas ($N_{\rm ants}$), scaling approximately as $\mathcal{O}(N_{\rm sources} N^{\alpha}_{\rm ants})$. For \fftvis{}, when utilizing its general Type 3 NUFFT, the scaling behavior reflects the combined costs of processing the sources and performing the NUFFT on its internal grid (whose size is influenced by $u_{\rm max}$). This leads to an approximate overall scaling of $\mathcal{O}(N_{\rm sources} + u^2_{\rm max} \log u^2_{\rm max})$. We use these scaling relations and the results of Figure \ref{fig:scaling} to project to expected runtime of a simulation mirroring our test case, but for modern 21\,cm arrays in Table \ref{tab:array_runtimes}.
 
These scaling results highlight the complementary strengths of \fftvis{} and \matvis{}. \fftvis{} offers substantial runtime gains for dense, many-element arrays with large sky models, scenarios in which the cost of NUFFT-based gridding and interpolation remains subdominant to the prohibitive scaling of per-antenna, per-source computations. In practice, the choice between simulators depends on the trade-off between sky model size, array density, and baseline length. Therefore, we recommend \fftvis{} for cases where the ratio of the number of antennas to the maximum extent of the array in $u$ is large (i.e., $N_{\rm ant} / u_{\rm max}$) or the array layout is gridded, and \matvis{} for simulation cases with low sky resolution and extended arrays. The benchmarking results presented here can help guide users in selecting the most appropriate simulator for their specific application.

\begin{table*}
  \centering
  \resizebox{\textwidth}{!}{%
    \begin{tabular}{|l|ccc|ccc|}
      \hline
       & 
       \multicolumn{3}{c|}{Diffuse Sky Model ($N_{\rm sources}=16\pi\,u^2_{\rm max}$)} 
       & \multicolumn{3}{c|}{Fixed Size Sky Model ($N_{\rm sources}=10^6$)} 
      \\ \hline
      \textbf{Array Layout} 
        & \multicolumn{1}{c|}{\texttt{matvis} (s)} 
        & \multicolumn{1}{c|}{\texttt{fftvis} (s)} 
        & \begin{tabular}[c]{@{}c@{}}Gridded \\ \texttt{fftvis} (s)\end{tabular} 
        & \multicolumn{1}{c|}{\texttt{matvis} (s)} 
        & \multicolumn{1}{c|}{\texttt{fftvis} (s)} 
        & \begin{tabular}[c]{@{}c@{}}Gridded \\ \texttt{fftvis} (s)\end{tabular} 
      \\ \hline
      HERA Core (150 MHz) 
        & \multicolumn{1}{c|}{$6.2\times10^2$} & \multicolumn{1}{c|}{$4.6$}   & $4.1$   & \multicolumn{1}{c|}{$6.25\times10^2$} & \multicolumn{1}{c|}{$4.3$}  & $3.84$  \\
      HERA-350 (150 MHz) 
        & \multicolumn{1}{c|}{$7.1\times10^3$} & \multicolumn{1}{c|}{$42.2$}  & $37.1$  & \multicolumn{1}{c|}{$6.9\times10^2$}  & \multicolumn{1}{c|}{$8.92$} & $3.84$  \\
      OVRO-LWA (50 MHz) 
        & \multicolumn{1}{c|}{$6.1\times10^3$} & \multicolumn{1}{c|}{$38.1$}  & --      & \multicolumn{1}{c|}{$6.9\times10^2$}  & \multicolumn{1}{c|}{$8.4$}  & --      \\
      SKA-LOW (200 MHz) 
        & \multicolumn{1}{c|}{$1.3\times10^8$} & \multicolumn{1}{c|}{$5.8\times10^5$} & -- & \multicolumn{1}{c|}{$1.04\times10^3$} & \multicolumn{1}{c|}{$1.14\times10^5$} & -- \\
      MWA Phase II Compact (175 MHz) 
        & \multicolumn{1}{c|}{$1.4\times10^3$} & \multicolumn{1}{c|}{$26.8$}   & --      & \multicolumn{1}{c|}{$2.3\times10^2$}  & \multicolumn{1}{c|}{$6.96$} & --      \\
      MWA Phase II Extended (175 MHz) 
        & \multicolumn{1}{c|}{$2.1\times10^5$} & \multicolumn{1}{c|}{$1.9\times10^3$} & -- 
        & \multicolumn{1}{c|}{$4.9\times10^2$}  & \multicolumn{1}{c|}{$2.9\times10^2$} & -- 
      \\ \hline
    \end{tabular}%
  }
  \caption{Comparison of the expected runtime of \texttt{fftvis} and \texttt{matvis} for several existing and planned 21\,cm arrays at one frequency at the center of the observing band for each instrument and 16 integrations. Here, we use the scaling relations determined in Figure \ref{fig:scaling} to determine the runtime, setting the number of sources such that pixel size is half the size of the angular resolution of set by the longest baseline in units of wavelengths. For comparison purposes, we also include a column for \texttt{matvis} and \texttt{fftvis} where we fix the sky model to $N_{\rm sources} = 10^6$ to demonstrate the scaling differences between \fftvis{} and \matvis{} with number of sources. We find that \fftvis{} outperforms \matvis{} for most modern arrays, except for SKA-Low when the sky model is fixed.}
  \label{tab:array_runtimes}
\end{table*}

\subsection{Memory Requirements}
\label{subsec:memory_usage}


A key improvement of \fftvis{} over \matvis{} is its substantially reduced memory footprint for dense, many-element arrays. This advantage arises from fundamental differences in how the two simulators evaluate the RIME. While \matvis{} constructs visibilities through explicit outer products across antennas, requiring the allocation of large $N_{\rm sources} N_{\rm ants}$ intermediate matrices, \fftvis{} uses NUFFTs, which reduces intermediate data volume and distributes memory costs more evenly across computation steps. As a result, \fftvis{} not only minimizes usage of memory per-node, but also allows for a more effective parallelization across shared-memory and distributed systems (see Section~\ref{subsec:multicore}).

The memory usage of \fftvis{} is typically dominated by a few large arrays, whose approximate scalings with relevant simulation parameters are summarized below. In these scalings, we define $N_{\rm sources}$ as the number of source components, $N_\nu$ as the number of frequency channels, $N_{\rm times}$ as the number of time integrations, $N_{\rm bls}$ as the number of baselines, $N_{\rm feed}$ as the number of feed polarizations, and $N_{\rm beam}$, $N_{\rm beampix}$ as the number of unique beam models (currently 1 for \fftvis{}) and the number of pixels in a stored beam map, respectively. The dimensions of the intermediate FFT grid required by the NUFFT in dimension $i$ are denoted by $n_i$.
\begin{enumerate}
    \item \textbf{Source Flux Array:} Stores frequency-dependent fluxes (or Stokes parameters) for all input sky components. Scales as $\mathcal{O}(N_{\rm sources} N_\nu)$.
    \item \textbf{Output Visibility Array:} Holds the final simulated visibilities. Scales as $\mathcal{O}(N_{\rm bls} N_{\nu} N_{\rm times} N_{\rm feed}^2)$.
    \item \textbf{Raw Beam Map:} Stores the input beam model if provided as a map (e.g., \texttt{UVBeam}). Scaling depends on the beam format, e.g., $\mathcal{O}(N_{\rm feed} N_\nu N_{\rm beampix} N_{\rm beam})$ for a frequency-dependent beam FITS file. If used, \texttt{AnalyticBeam} objects have negligible storage cost.
    \item \textbf{Interpolated Beam Values:} Stores beam values evaluated at source locations for the current time step. Scales as $\mathcal{O}(N_{\rm feed} N_{\rm sources} N_{\rm beam})$.
    \item \textbf{Source Coordinates:} Stores source coordinates. Scales as $\mathcal{O}(N_{\rm sources})$.
    \item \textbf{NUFFT Intermediate Grid:} The temporary uniform grid used internally by \texttt{finufft}. Its size in each dimension $i$, $n_i$, is determined by Equation~\ref{eqn:grid_size}, which depends on the maximum source extent (which we take to be $S_i = 1$ for sky models that extend to the horizon), the maximum baseline coordinate, $X_i$, the desired precision, $\epsilon$, and oversampling factor, $\sigma$. The memory explicitly scales as $4 \sigma N_{\rm grid}$.
\end{enumerate}

\begin{figure*}
    \centering    \includegraphics[width=\textwidth]{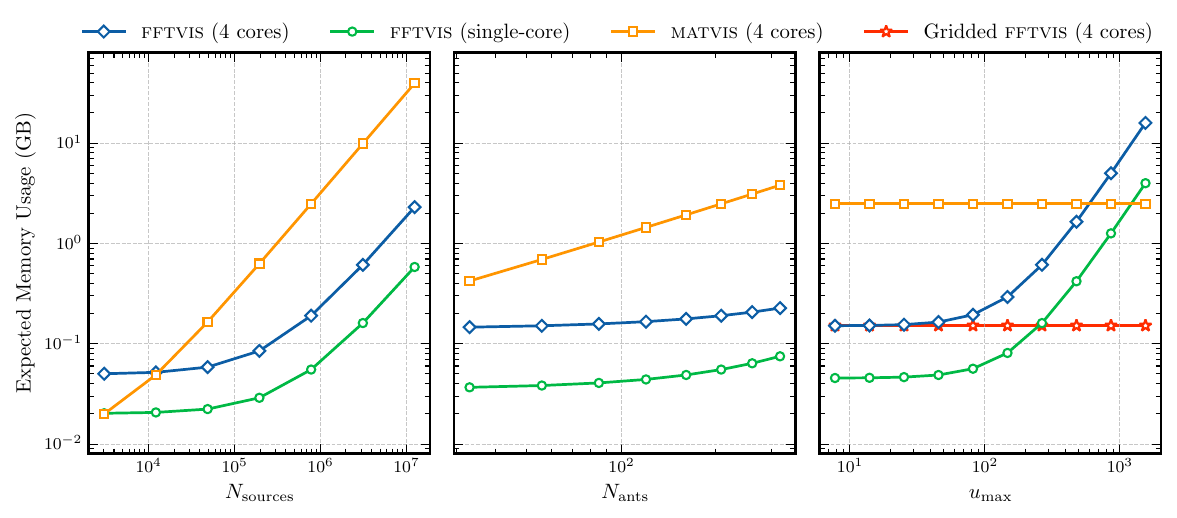}
    \caption{Comparison of the estimated memory usage of \fftvis{} and \matvis{} for the computational scaling cases presented in \Cref{fig:scaling}. For results shown here, we assume no chunking of the sky model in the processing stage, which can reduce the memory usage for \matvis{} by summing over subsets of the sky model. We find that memory usage of \fftvis{} is significantly less than that of \matvis{} especially when the number of antennas and sources is large due to \matvis{} forming an $N_{\rm sources} N_{\rm ants}$ matrix. When $u_{\max}$ is large, \fftvis{} memory usage exceeds that of \matvis{} due to the formation of the underlying fine FFT grid that scales with the physical size of the array in units of wavelengths. As with the computational scaling explored in Figure \ref{fig:scaling}, the memory usage for arrays with a large $u_{\rm max}$ can be dramatically reduced if the array is gridded, as shown at in the right-hand column.}
\label{fig:memory_scaling}
\end{figure*}

In Figure~\ref{fig:memory_scaling}, we present projected memory estimates for both \fftvis{} and \matvis{} across the same range of array and sky model sizes explored in Figure~\ref{fig:scaling}. The left and middle panels show that for increasing numbers of sources ($N_{\rm sources}$) and antennas ($N_{\rm ants}$), \fftvis{} has a significantly reduced memory usage than \matvis, often by more than an order of magnitude. This reflects the cost of forming and storing the outer product matrices in \matvis{}, which scales unfavorably for large $N_{\rm ants}$ and $N_{\rm sources}$. The right panel shows a contrasting case in which \fftvis{} incurs a higher memory load. In particular, simulations with large maximum baseline lengths increase the size of the oversampled FFT grid internal to \finufft{} that dominates memory usage. These grid dimensions scale linearly with $u_{\rm max}$. However, when the input array can be defined on a regular grid and a Type 1 transform can be used to evaluate the RIME, memory usage can be dramatically reduced. In this case, the size of the grid used for the FFT is only as large as the extent of the baseline vectors in the rotated coordinate space. This grid size can potentially be substantially smaller than the intermediate grid used by the Type 3 transform for arrays with large $u_{\rm max}$ as shown in the right-hand column of Figure \ref{fig:memory_scaling}.

In \matvis{}, a practical approach to managing memory involves chunking computations along the source axis. This technique allows for a nearly arbitrary reduction in the peak memory usage by processing subsets of sources sequentially, with generally minimal computational overhead. While the overhead might increase for very fine-grained chunking, it typically remains manageable for current hardware memory capacities. However, this source chunking strategy is generally less useful for \fftvis{} when the maximum extent of the array is large. As shown in the memory tests above, the memory usage of \fftvis{} is typically dominated by the size of the intermediate FFT grid. This grid size is not dependent on the number of sources, but rather on the maximum angular extent of the sky model being considered ($S_i$) and the maximum baseline coordinates ($X_i$). As a result, simply reducing the number of sources processed per chunk does not shrink the required grid size if the sources across all chunks collectively span the same wide angular extent. One potential way to mitigate poor memory scaling in \fftvis{} would involve spatial chunking, or dividing the simulations into chunks based on the sources occupying localized regions of the sky. By limiting the angular extent ($S_i$) for sources within a single chunk, the factors determining the grid size in Equation \ref{eqn:grid_size} could be reduced, thus decreasing the maximum memory required for processing that chunk. Implementing such a scheme would require careful partitioning of the source catalog and potentially reinitializing the NUFFT for each spatial chunk, which may increase the computational runtime. We leave the development and evaluation of this spatial chunking method for future work.

Overall, \fftvis{} demonstrates superior memory scaling for simulation configurations relevant to densely-packed 21\,cm experiments, especially for arrays that can be defined on a gridded layout. By minimizing memory usage and avoiding bottlenecks associated with matrix expansion, it enables large-scale simulations on lower memory machines and improves efficient parallelism across distributed compute environments.

\subsection{Multicore Performance and Parallelization}
\label{subsec:multicore}
Ensuring that our software scales efficiently with available hardware is essential for maximizing the performance of \fftvis. As modern computing architectures increasingly feature high core counts, it is essential to design software that effectively distributes workloads across multiple cores. There are two primary approaches to leveraging multi-core systems: (1) utilizing multi-threading within individual algorithmic components to take advantage of shared-memory parallelism, and (2) explicitly dividing computations into independent tasks that run across multiple processes.

In \fftvis, we adopt a hybrid strategy that prioritizes multiprocessing while still incorporating multithreading. Specifically, we parallelize the entire workflow across multiple processes, then assign the remaining cores as threads within each process. This approach is motivated by findings in \cite{2025RASTI...4....1K}, where a pure multithreaded strategy exhibited a performance plateau after just a few cores due to inherent scaling limitations and shared resource contention. In contrast, a multiprocessing strategy avoids many of these bottlenecks, providing more favorable scaling across a larger number of cores, albeit with a modest overhead associated with initializing processes.

To manage the distribution of these independent tasks across processes on a single machine, \fftvis{} takes advantage of the \texttt{ray} library. Originally developed for scalable machine learning, \texttt{ray} provides a robust framework for efficiently scaling Python applications from a single core to many. Specifically, \fftvis{} uses \texttt{ray} to divide the overall simulation workload by distributing the calculations across the $N_{\rm proc}$ available processes, assigning each process a roughly equal portion of the total workload determined by the number of simulation time steps and frequency channels. This approach helps manage the creation and coordination of multiple processes, simplifying the implementation of our chosen multiprocessing strategy. Furthermore, \texttt{ray} supports shared memory, enabling large, read-only objects like the beam, sky model, and model coordinates to be loaded once and accessed by multiple worker processes. 

The total memory requirement of \texttt{fftvis} in this shared memory state can be approximated as
\begin{equation}
M_{\mathrm{tot}} \approx M_{\mathrm{shared}} + N_{\mathrm{proc}}\,M_{\mathrm{local}},
\label{eq:mem_scaling}
\end{equation}
where $M_{\mathrm{shared}}$ represents arrays that can reside in shared memory 
(e.g., raw beam values and sky model components), and $M_{\mathrm{local}}$ denotes 
per-process allocations such as interpolated beam grids and per-source coordinate 
transformations. Explicitly,
\begin{align}
M_{\mathrm{shared}} &\simeq \underbrace{N_{\mathrm{sources}}\,N_{\nu}\,\beta_{\mathrm{f}}}_{\text{source flux}} +
\underbrace{N_{\rm bls}\,
N_{\nu}\,N_{\mathrm{feed}}^{2}\,N_{\rm times}\,\beta_{\mathrm{c}}}_{\text{visibility array}} \nonumber \\ 
& \ \ + \underbrace{N_{\nu}\,N_{\mathrm{beampix}}\,\beta_{\mathrm{c}}}_{\text{raw beam storage}},
\label{eq:mshared}
\end{align}
\begin{align}
M_{\mathrm{local}} &\simeq 
\underbrace{N^2_{\mathrm{feed}}\,N_{\mathrm{sources}}\,\beta_{\mathrm{c}}}_{\text{interpolated beam}} + \underbrace{3\,N_{\mathrm{sources}}\,\beta_{\mathrm{f}}}_{\text{source coordinates}}
+ \underbrace{N_{\mathrm{feed}}^{2}\,n_{x}\,n_{y}\,\beta_{\mathrm{c}}}_{\text{NUFFT grid}},
\label{eq:mlocal}
\end{align}
where $\beta_{\mathrm{f}}$ and $\beta_{\mathrm{c}}$ 
are the byte sizes of real and complex double-precision values, respectively. 
In the current implementation, $M_{\mathrm{shared}}$ includes large 
arrays that are loaded once and accessed by all worker processes, whereas 
$M_{\mathrm{local}}$ scales linearly with $N_{\mathrm{proc}}$,  
increasing the total memory with the number of processes. While this shared memory capability is beneficial, a current limitation within \fftvis's implementation is that each process still independently interpolates the beam and performs coordinate rotations for all $N_{\rm sources}$ on each process independently. Therefore, these specific steps do not fully utilize memory sharing, requiring that arrays of size $\mathcal{O}\left(N_{\rm sources}\right)$ be stored in memory for each independent process.

\begin{figure}
    \centering    \includegraphics[width=\columnwidth]{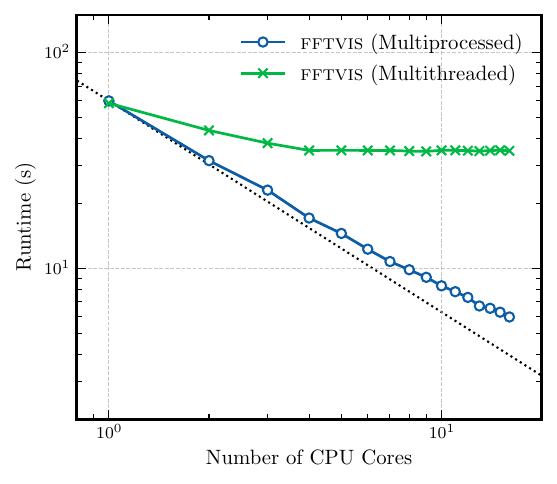}
    \caption{Runtime of \fftvis{} as a function of the number of CPU cores, demonstrating how its out-of-the-box performance improves with parallelization. Each curve underlaid with a dotted line representing ideal linear scaling, which serves as a benchmark for perfect parallel efficiency. As the number of available cores increases, \fftvis{} continues to scale efficiently, achieving near-optimal performance due to its explicit parallelization of each step in the algorithm. Deviation from ideal scaling results from a small amount of near constant overhead when initializing the independent processes, which becomes less significant for larger jobs.}
    \label{fig:nproc_scaling}
\end{figure}

We evaluated the scaling behavior of \fftvis{} using a benchmark simulation setup similar to that described in Section \ref{subsec:runtime}, but extended the total number of integrations to 256 to reduce the fractional impact of initialization in the timing. The number of processes was varied from 1 to 16, and execution times were measured using the \texttt{line\_profiler} library. We present these timings in Figure \ref{fig:nproc_scaling}. Here, we isolate the effects of multiprocessing or multithreading by controlling the other factor using the \texttt{threadpoolctl} library, thus demonstrating the scaling potential of each strategy independently. The results presented in Figure \ref{fig:nproc_scaling} confirm that \fftvis{} scales significantly better when tasks are distributed across multiple processes compared to relying solely on multithreading. This validates our hybrid approach, demonstrating that the performance gains outweigh the process initialization overhead.

To balance performance and memory constraints, we recommend configuring \fftvis{} such that total memory usage remains within the available system limits. A practical guideline is to set the number of threads per process as \begin{equation} 
N_{\rm threads} = \left\lfloor \frac{N_{\rm cores}}{N_{\rm proc}} \right\rfloor, 
\label{eqn:nproc}
\end{equation} which ensures that both multiprocessing and multithreading are utilized optimally. By default, \fftvis{} selects $N_{\rm threads}$ using Equation \ref{eqn:nproc} given a value of $N_{\rm proc}$ provided by the user, but does not enforce a value $N_{\rm proc}$ that keeps the simulator within system memory requirements. Therefore, it is important to estimate the total memory usage of \fftvis{} using Section \ref{subsec:memory_usage} as a guide, and select $N_{\rm proc}$ given the system constraints. By carefully tuning $N_{\rm proc}$ and $N_{\rm threads}$, users can minimize overhead while achieving an efficient and scalable execution of \fftvis.

While the performance analysis presented here is restricted to a single computational node, the underlying \fftvis{} framework is readily extensible to multi-node environments.  In our current deployment on NMASCO and similar clusters, distributed execution is achieved by partitioning the total bandwidth into independent frequency blocks and submitting these as separate \texttt{slurm} array jobs, each running an instance of \fftvis{} with shared configuration files.  Because visibilities at different frequencies are computed independently, this strategy yields near-ideal scaling across nodes with a small amount of communication overhead.  More generally, since \fftvis{} is built on the \texttt{ray} distributed execution framework, native multi-node parallelism can be supported directly within the package.  Future development will focus on integrating this capability into the standard workflow, allowing \texttt{ray} to manage task scheduling and inter-node communication transparently, streamlining large-scale simulations without reliance on external job arrays.

\section{Conclusion} \label{sec:conclusion}

The advancement of 21\,cm cosmology relies heavily on accurate and efficient simulations to validate analysis pipelines and understand instrumental systematics, a task often challenged by the computational expense of traditional visibility simulation methods. To address this challenge, we introduced \fftvis{}, a high-performance interferometric visibility simulator designed to accelerate forward modeling by using the Non-Uniform Fast Fourier Transform (NUFFT). By utilizing the \texttt{finufft} library, \fftvis{} efficiently evaluates the radio interferometry measurement equation (RIME) under the point-source approximation, leading to significant advantages in speed and memory efficiency compared to direct-summation techniques, particularly for simulations involving densely-packed interferometric arrays and sky models with millions of sources.

We have demonstrated through a series of validation tests that \fftvis{} achieves numerical accuracy comparable to the established visibility simulator, \matvis{}, itself validated against \texttt{pyuvsim} in \citet{2025RASTI...4....1K}. Our first set of comparisons against analytic solutions for diffuse sky models confirms the accuracy of the NUFFT implementation, while assessments of delay and fringe-rate spectra show that \fftvis{} preserves the spectral and temporal coherence required for end-to-end validation tests. We find that the visibilities produced do not introduce significant spectral or temporal artifacts when operated at sufficient NUFFT precision ($\epsilon \lesssim 10^{-13}$). Furthermore, our benchmarking tests highlight the specific regimes where \fftvis{} offers substantial advantages. For simulations involving large, densely-packed arrays (large $N_{\rm ant} / \sqrt{u_{\rm max}}$), \fftvis{} delivers runtime speedups of up to two orders of magnitude and significantly reduced memory usage compared to \matvis{}.

However, our performance tests also identified scenarios where the NUFFT approach is less efficient. For sparse arrays with large maximum baseline lengths ($u_{\rm max}$), the computational cost associated with the large intermediate FFT grid required by \fftvis{} can exceed that of direct summation methods like \matvis{}. We find that this limitation is significantly mitigated when simulating arrays that conform to a regular grid layout. In these instances, \fftvis{} can utilize a more efficient Type 1 NUFFT, substantially reducing the computational scaling and memory requirements especially for physically large arrays. More broadly, these performance differences between the two simulators highlight \fftvis{} as a powerful complement to existing simulators such as \matvis, offering improved simulation performance for the specific case of simulating large-$N$, densely-packed arrays typical of 21\,cm experiments like HERA and OVRO-LWA.

Throughout this work, we have identified several areas of future development that promise to further enhance the flexibility of \fftvis. One straightforward addition to the simulator will be to extend \fftvis{} to use GPU acceleration via \texttt{cufinufft} \citep{shih2021cufinufftloadbalancedgpulibrary} and the existing GPU components that exist in the \matvis{} codebase. This will potentially improve the performance of \fftvis{} by an additional order-of-magnitude, making large-scale simulations even more tractable. Another potential area of improvement is removing the current limitation of supporting only a single, array-wide beam model by implementing per-antenna beam simulations. As discussed in Section \ref{subsubsec:fftvis_beams}, \fftvis{} is expected to scale poorly with the number of unique beams requested by a user. However, this extension would offer the enhanced flexibility necessary for some systematic studies, and may remain more efficient than \matvis{} if the number of unique beams is small. Finally, we also plan on exploring advanced techniques, such as the spectral basis decomposition outlined in Appendix~\ref{appendix:basis_decomp} to reduce NUFFT calls for wide-band simulations of spectrally smooth sources and beams, which may also improve \fftvis{} simulation efficiency.

With these improvements, \fftvis{} is positioned to become an increasingly powerful tool for the low-frequency radio astronomy community. We anticipate that its demonstrated combination of speed, accuracy, and scalability, coupled with future improvements, will significantly benefit ongoing and future 21\,cm cosmology experiments by enabling efficient end-to-end pipeline validation, systematics exploration, and forward-modeling studies. The \fftvis{} package is publicly available and open-source, and we encourage community involvement in its continued development.

\section*{Acknowledgements}
This material is based upon work supported by the National Science Foundation under grants \#1636646 and \#1836019 and institutional support from the HERA collaboration partners. This research is funded in part by the Gordon and Betty Moore Foundation
through Grant GBMF5212 to the Massachusetts Institute of Technology. HERA is hosted by the South African Radio Astronomy Observatory, which is a facility of the National Research Foundation, an agency of the Department of Science and Innovation.
This project has received funding from the European Union’s Horizon 2020 research and innovation programme under the Marie Skłodowska-Curie grant agreement No 101067043.

\section*{Data Availability}

All timing benchmarks and simulated visibilities in this work were generated with the publicly‑available \texttt{fftvis} package (v1.0.0). The \texttt{fftvis} source is hosted on GitHub at \url{https://github.com/tyler-a-cox/fftvis}. The analysis and simulation scripts (parameter files, run‑scripts, etc.) used to produce the exact results in Figures 2--8 are available from the corresponding author upon reasonable request.

\section*{Conflict of Interest}
Authors declare no conflict of interest.

\section*{Code Availability}
All components of the \fftvis{} package\footnote{\url{https://github.com/tyler-a-cox/fftvis}} and the associated software mentioned in this paper, including \texttt{hera\_sim}\footnote{\url{https://github.com/HERA-team/hera_sim}}, \texttt{pyuvsim}\footnote{\url{https://github.com/RadioAstronomySoftwareGroup/pyuvsim}}, and \texttt{pyuvdata} \footnote{\url{https://github.com/RadioAstronomySoftwareGroup/pyuvdata}} are publicly available under open source licenses on \texttt{GitHub}. Distributions are also available via PyPI distributions, allowing easy installation via the \texttt{pip install} command. The codebase adheres to best practices in collaborative development, including continuous integration, test coverage reporting, and extensive API documentation. Example notebooks and tutorials are provided as part of the repository to guide new users through basic and advanced usage scenarios. Although the primary contributions to the \fftvis{} code have so far come from the authors of this paper, we welcome and encourage contributions from the broader community.

\section*{Software}
The work in this paper was made possible by the following software packages: \texttt{numpy} \citep{harris2020array}, \texttt{pyuvdata} \citep{2017JOSS....2..140H}, \texttt{scipy} \citep{2020SciPy-NMeth}, \texttt{matplotlib} \citep{Hunter:2007}, \texttt{finufft} \citep{barnett2019parallelnonuniformfastfourier}, \texttt{astropy} \citep{2013A&A...558A..33A, 2018AJ....156..123A}, \texttt{healpy} \citep{2019JOSS....4.1298Z}, and \texttt{pygdsm} \citep{2016ascl.soft03013P}

\bibliographystyle{mnras}
\bibliography{references, software}

\appendix
\section{Efficient Evaluation of Spectrally Smooth Sky Models with \fftvis{}}
\label{appendix:basis_decomp}

For simulations involving many frequency channels, particularly when the sky model and beam exhibit smooth spectral variations, significant improvements in the computational efficiency can be achieved in \fftvis{} by representing these inputs using a compact spectral basis. This approach capitalizes on the inherent spectral smoothness, enabling the frequency dependence to be accurately represented by a small number of basis functions.
 
Prior to performing source rotation and beam interpolation, the sky model, $\mathbf{s}$, and beam, $\mathbf{b}$, are decomposed into a chosen compact basis ($\mathbf{M}$) that efficiently captures their smooth spectral evolution. The decomposition is achieved through a least-squares projection onto the basis for each pixel in the sky and beam models,
\begin{align}
\hat{\mathbf{s}} &= \left(\mathbf{M}^{\rm T}\mathbf{M}\right)^{-1} \mathbf{M}^{\rm T} \mathbf{s} \\
\hat{\mathbf{b}} &= \left(\mathbf{M}^{\rm T}\mathbf{M}\right)^{-1} \mathbf{M}^{\rm T} \mathbf{b}.
\end{align}
Here, $\hat{\mathbf{s}}$ and $\hat{\mathbf{b}}$ denote the precomputed basis coefficients for the sky and beam, respectively. These coefficients are stored for use at each subsequent time step.

During each time step, the stored beam coefficients are interpolated to the instantaneous positions of the sky model sources. The visibility for each source is then computed by efficiently combining the precomputed basis coefficients for the sky and beam. Instead of directly multiplying their full frequency-space representations, we exploit the multiplicative properties inherent in the basis expansion. This product is represented compactly as:
\begin{equation}
\mathbf{v}_k = \sum_{i, j} \mathbf{s}_{i} \mathbf{b}_{j} \mathbf{d}^k_{ij},
\end{equation}
where $\mathbf{d}^k_{ij}$ are the structure constants of the basis $\mathbf{M}$, defined as
\begin{equation}
    \mathbf{d}_{ij}^k = \sum_q \mathbf{M}_{qi} \mathbf{M}_{qj} \mathbf{M}_{qk}.
\end{equation}
These constants reflect how the product of two basis functions can itself be expressed within the same basis, thus avoiding a computationally expensive reprojection. Once the product coefficients $\mathbf{v}_k$ are computed, a type-2 NUFFT (non-uniform to uniform) transforms this representation from a non-uniform to a uniform grid in the $uv$-plane. The resolution of this uniform grid is carefully chosen to satisfy the resolution constraints imposed by the highest observing frequency, defined as:
\begin{equation}
N{i} = \frac{2 \sigma}{\pi c} b_{\max} \nu_{\max},
\end{equation}
where $\sigma$ is a scaling parameter, $b_{\max}$ is the maximum baseline length, and $\nu_{\max}$ is the maximum observing frequency.

Finally, the uniform $uv$-plane representation obtained from the NUFFT for each of the basis functions is expanded to frequency space and interpolated to the user-specified baseline positions. By expanding the compact basis representation back into the full frequency domain, the simulator effectively generates the required visibilities without repetitive frequency-dependent gridding and Fourier transformations. This compact basis representation substantially reduces computational load on the NUFFT by encoding the smooth spectral information of the sky and beam into a small set of efficiently managed coefficients. Overall, the efficient product computation via structure constants, and the strategic use of the NUFFT should allow \fftvis{} to more efficiently handle simulations across numerous frequency channels when the NUFFT dominates the runtime of the simulation.

\bsp	
\label{lastpage}
\end{document}